\documentclass[12pt]{article}

\RequirePackage[OT1]{fontenc}
\usepackage{natbib}
\RequirePackage[colorlinks,citecolor=blue,urlcolor=blue]{hyperref}
\usepackage[english]{babel}
\usepackage{amsthm}
\usepackage{amsmath,bigints}
\usepackage{amsfonts}
\usepackage{amssymb}
\usepackage{graphicx,subcaption}
\usepackage{enumerate}
\usepackage{color}
\usepackage{array}
\usepackage{psfrag,epsf}
\usepackage{url} % not crucial - just used below for the URL 
\usepackage[export]{adjustbox}
\usepackage{bm,float}
\usepackage{titling}

\def\frac#1#2{{\textstyle{#1\over#2}}}

\DeclareSymbolFont{AMSb}{U}{msb}{m}{n}
\DeclareMathSymbol{\Natural}{\mathbin}{AMSb}{"4E}
\DeclareMathSymbol{\Integer}{\mathbin}{AMSb}{"5A}
\DeclareMathSymbol{\Real}{\mathbin}{AMSb}{"52}
\DeclareMathSymbol{\Rational}{\mathbin}{AMSb}{"51}
\DeclareMathSymbol{\Imaginary}{\mathbin}{AMSb}{"49}
\DeclareMathSymbol{\Complex}{\mathbin}{AMSb}{"43} 
\DeclareMathSymbol{\Disk}{\mathbin}{AMSb}{"44} 
\def\bi{\begin{itemize}}
\def\ei{\end{itemize}}
\def\bd{\begin{description}}
\def\ed{\end{description}}
\def\ben{\begin{enumerate}}
\def\een{\end{enumerate}}
\def\bv{\begin{verbatim}}
\def\ev{\end{verbatim}}

\def\hat#1{{\widehat{#1}}}

\def\2to{{\ {\buildrel 2\over \longrightarrow}\ }}

\def\I1ton{{$I_1,\ldots,I_n$}}
\def\X1ton{{$X_1,\ldots,X_n$}}
\def\Y1ton{{$Y_1,\ldots,Y_n$}}
\def\Z1ton{{$Z_1,\ldots,Z_n$}}
\def\R1ton{{$R_1,\ldots,R_n$}}
\def\e1ton{{$e_1,\ldots,e_n$}}
\def\t1ton{{$t_1,\ldots,t_n$}}
\def\x1ton{{$x_1,\ldots,x_n$}}
\def\y1ton{{$y_1,\ldots,y_n$}}
\def\z1ton{{$z_1,\ldots,z_n$}}
\begin{document}
\thispagestyle{empty}
\baselineskip=28pt
\vskip 5mm
\begin{center} {\Large{\textbf{Approximate Bayesian inference for high-resolution spatial disaggregation using alternative data sources}}}
\end{center}

% Approximate Bayesian Inference for 
% A sparse Gaussian scale mixture process for short-range extremal dependence and long-range independence 

%estimating

\baselineskip=12pt
\vskip 5mm

\begin{center}
\large
Anis Pakrashi$^{1,2}$, Arnab Hazra$^{2}$, \\ Sooraj M Raveendran$^{3}$, and Krishnachandran Balakrishnan$^{3}$
\end{center}

\footnotetext[1]{
\baselineskip=10pt 
Department of Statistics, Pennsylvania State University, University Park, Pennsylvania 16802, USA. \\ E-mail: avp6357@psu.edu}
\footnotetext[2]{
\baselineskip=10pt Department of Mathematics and Statistics, Indian Institute of Technology Kanpur, Kanpur 208016, India. \\ E-mail: ahazra@iitk.ac.in}
\footnotetext[3]{
\baselineskip=10pt Indian Institute of Human Settlements, Armane Nagar, Bengaluru, Karnataka 560080, India. \\ E-mail: soorajmr@iihs.ac.in, krishna@mapsolve.ai}
% \footnotetext[2]{
% \baselineskip=10pt Insert affiliation of second author}
% \footnotetext[3]{
% \baselineskip=10pt Insert affiliation of third author}

\baselineskip=17pt
\vskip 4mm
\centerline{\today}
\vskip 6mm

%%%%%%%%%%%%%%%%%%%%%%%%%%%%%%%%%%%%%%%%%%%%%%%%%%%%%%%%%%%%%%%%%%%%%%%% like a 30m $\times$ 30m pixel
\begin{center}
{\large{\bf Abstract}}
\end{center}
This paper addresses the challenge of obtaining precise demographic information at a fine-grained spatial level, a necessity for planning localized public services such as water distribution networks, or understanding local human impacts on the ecosystem. While population sizes are commonly available for large administrative areas, such as wards in India, practical applications often demand knowledge of population density at smaller spatial scales. We explore the integration of alternative data sources, specifically satellite-derived products, including land cover, land use, street density, building heights, vegetation coverage, and drainage density. Using a case study focused on Bangalore City, India, with a ward-level population dataset for 198 wards and satellite-derived sources covering 786,702 pixels at a resolution of 30m×30m, we propose a semiparametric Bayesian spatial regression model for obtaining pixel-level population estimates. Given the high dimensionality of the problem, exact Bayesian inference is deemed impractical; we discuss an approximate Bayesian inference scheme based on the recently proposed max-and-smooth approach, a combination of Laplace approximation and Markov chain Monte Carlo. A simulation study validates the reasonable performance of our inferential approach. Mapping pixel-level estimates to the ward level demonstrates the effectiveness of our method in capturing the spatial distribution of population sizes. While our case study focuses on a demographic application, the methodology developed here readily applies to count-type spatial datasets from various scientific disciplines, where high-resolution alternative data sources are available.

% The model incorporates a Gaussian process to represent the underlying (transformed) intensity function and accounts for the effects of alternative data sources. 
% potential nonlinear 

\baselineskip=16pt

\par\vfill\noindent

\pagenumbering{arabic}
\baselineskip=24pt

\newpage

% collection about detailed processes using

%\newpage
%\spacingset{1.45} % DON'T change the spacing!
\section{Introduction}
\label{sec:intro}

%Spatially-referenced variables often exhibit inherent correlation among locations, where observations from nearby locations demonstrate more similarity compared to those from distant locations \citep{cressie1993statistics}. 

Practical social problems often require analyzing data at finer-resolution spatial regions. The process of transitioning data from a higher (or finer) to a lower (or coarser) resolution is known as aggregation \citep{roquette2018relevance,paige2022spatial}. The delineation of boundaries frequently depends on the specific problem at hand, particularly observed in settings involving census data. However, utilizing statistical models based on aggregated data introduces certain natural disadvantages \citep{pollet2015taking}. In various research domains such as forestry, agronomy, meteorology, public health, epidemiology, and soil science, information aggregation poses a significant challenge \citep{van1999aggregation,rudstrom2002data}. Finer-resolution unit information is often obscured due to aggregating data, making intricate trends invisible. The limitations associated with models for aggregated data drive the need for methodologies to recover the original (pixel-level) information from coarser resolution observations. This reverse process is referred to as spatial downscaling or disaggregation, with examples of applications provided by \cite{mertens1997spatial} and \cite{muhling2018potential}. Spatial disaggregation finds applications in hydrology \citep{ALBER2011343}, census data \citep{ijgi8080327}, climate \citep{segond2007simulation}, agriculture \citep{you2009generating}, disease study \citep{arambepola2022simulation}, health \citep{utazi2019spatial}, and other fields.

% Although the dataset comprises a total of three response variables, our emphasis is exclusively on population counts. 

Our study is motivated by a practical issue concerning data availability at larger spatial units, such as districts or wards in India, which commonly represent administrative boundaries. However, for applications requiring the subdivision of an entire city into distinct water-supply zones or understanding local human impacts on the ecosystem, ward-level data may not accurately depict the spatial distribution of the population, as noted by \cite{sawicki1973studies}. Addressing this challenge involves spatial disaggregation, aiming to predict population figures at smaller spatial units, such as 30m $\times$ 30m pixels, which can then be aggregated based on specific requirements. Our study addresses the spatial disaggregation problem using a comprehensive dataset sourced from the Indian Institute of Human Settlements (IIHS), a national institution dedicated to the advancement and transformation of Indian cities and settlements. The primary objective is to employ covariates at a resolution of 30m $\times$ 30m cells for disaggregating the population of Bangalore city, as documented by \cite{sudhira2007city}. We focus particularly on Bruhat Bengaluru Mahanagara Palike (BBMP), the administrative body for the Bangalore metropolitan area, utilizing the 2011 Census data divided into 198 wards. Additional details about the setup and background can be found in \cite{balakrishnan2020method}. A possible solution is conducting high-resolution spatial disaggregation, drawing insights from relevant studies \citep{earnest2010small,utazi2019spatial,sadik2020small}. A homogeneous Poisson Process assumption is clearly unsuitable. A spatially-varying intensity function of the underlying nonhomogeneous Poisson process can be well estimated using a semiparametric approach or a spline-based method. A Gaussian process approximation to the likelihood is also effective due to the large population sizes in each ward. Additional applications of the Gaussian process in spatial analysis are in \cite{tapia2016prediction} and \cite{bullock2023latent}.

% because each ward has different population densities
% The fact that the population can follow a non-homogeneous Poisson Process can be a suitable assumption because each ward has different population densities. 

Alternative data sources are indispensable in statistical research, particularly in spatial data studies, due to their capacity to complement conventional datasets, address data gaps, enhance precision, facilitate exploration of novel research questions, and increase robustness through data integration. These alternative sources, encompassing remote sensing data, social media geotagged information, and crowd-sourced data, offer distinct advantages including real-time insights, finer resolution, accessibility to remote regions, and the facilitation of novel methodologies and research domains \citep{foulkes2008using, golder2011diurnal, wulder2012landsat, machado2021alternative, de2023collaborative}. By harnessing the diversity of these data sources, researchers can surmount the limitations inherent in traditional datasets, thus revealing deeper insights into multifaceted phenomena spanning various disciplines. In our study, data for several important predictors of population density were collected using satellite imaging, deep learning, and computer vision techniques, resulting in datasets with high-resolution information essential for estimating responses at finer spatial resolutions.

Different approaches in the past have led to a vast repository of disaggregation approaches, although each has its own limitations.
\cite{mertens1997spatial} discuss an application to deforestation modeling using traditional spatial modeling techniques. \cite{muhling2018potential} explain disaggregation approaches like bias-corrected quantile mapping (BCQM), change factor quantile mapping (CFQM), equidistant quantile mapping (EDQM), and the cumulative distribution function transform (CDFt) on various computer models for water temperature and salinity, called general circulation models. \cite{utazi2019spatial} develop a methodology for high-resolution mapping of vaccination coverage using a binomial spatial regression model with a logit link and a combination of covariate data and random effects modeling two levels of spatial autocorrelation in the linear predictor. The authors build their proposed Bayesian model using a stochastic partial differential equation (SPDE) approach of \cite{lindgren2011explicit} and the computation involves the integrated nested Laplace approximation (INLA) approach of \cite{rue2009approximate}. \cite{JSSv106i11} design an \texttt{R} package \texttt{disaggregation} to implement spatial disaggregation; they also use SPDE, an approximation to dense Gaussian processes (GPs) using Gaussian Markov random fields that allow sparse precision matrices. The approximation errors are unavoidable and depend on the mesh construction and the true range of the underlying GP which is generally unknown for real datasets \citep{hazra2021realistic, cisneros2023combined}. Thus, a true dense GP, if computationally feasible, is preferred. Besides, a very high-resolution (30m $\times$ 30m, for example) spatial disaggregation using these above approaches can be highly computationally challenging. %Besides, the illustrations in \cite{JSSv106i11} use 

 % In this paper, a simpler approach using Bayesian models have been used for the same.
 
Bayesian hierarchical models have proven instrumental in simplifying intricate Bayesian problems, effectively addressing problems characterized by unknown or complex joint distributions by breaking them down into multilevel structures with priors and hyperpriors at various levels \citep{schmid2000bayesian}. By accommodating multiple sources of variability, hierarchical models provide consistent and accurate estimates, demonstrating broad applicability across various real-life domains, including medical sciences \citep{yang2022classification,li2023combined}, climatology and geosciences \citep{berliner2000long, wainwright2016hierarchical}, as well as in ecology \citep{wikle2003hierarchical,ponciano2009hierarchical}. In spatial disaggregation or downscaling problems, these models have been found particularly useful \citep{anjoy2019estimation, shiferaw2023mapping}. Using a posterior predictive approach, fine-scale information can be obtained by employing a multiple-layer Bayesian structure, each corresponding to an aspect at a coarser resolution \citep{tassone2010disaggregated, tasic2016applications, irekponor2022framework, murphy2023bayesian}. Bayesian latent Gaussian models are Bayesian hierarchical models that assign Gaussian prior densities to the latent parameters and they are widely used in different scientific disciplines \citep{Hazra2023, Hrafnkelsson2023}.

%Gibbs sampling with closed-form full conditional posteriors is a good way to tackle Bayesian inference provided we can ensure a conjugate prior for every model parameter \citep{gelfand2000gibbs}.

% \newpage

% Approximate Bayesian inference

Closed-form expressions of the posterior distributions of the model parameters exist mostly in naive examples and the posteriors usually involve high-dimensional integrals. Techniques such as variational inference, Markov chain Monte Carlo (MCMC), and sequential Monte Carlo (SMC) provide avenues for approximating posterior distributions, circumventing the need for explicit computation of high-dimensional integrals \citep{brooks2011handbook}. These methods find broad application across diverse fields, including machine learning, computational biology, econometrics, and computational neuroscience, facilitating probabilistic modeling, uncertainty quantification, parameter estimation, and forecasting \citep{harva2008algorithms, martino2011approximate, stumpf2014approximate}. For Bayesian latent Gaussian spatial models, standard MCMC algorithms are commonly used \citep{hazra2021realistic} and INLA is a common deterministic computation-based algorithm \citep{utazi2019spatial}. In case the likelihood and prior both are Gaussian, the posterior is also Gaussian due to conjugacy, and several computationally attractive tools are available in the literature for sampling from high-dimensional Gaussian posteriors \citep{gelfand2016spatial}. However, in the likelihood is non-Gaussian, such a conjugacy does not hold; using a random-walk Metropolis-Hastings algorithm or its advanced variants for updating parameters is common here \citep{yadav2023joint}. When the dimension of the parameter vector is extremely high (786,702 in our case), existing exact computing tools are not feasible and an approximate Bayesian inference is a possible solution here. A recently proposed Max-and-Smooth approach \citep{hrafnkelsson2021max} approximates the non-Gaussian likelihood using a Gaussian likelihood via Laplace approximation and subsequently, for a Gaussian prior, the conjugacy of the prior, and the approximated likelihood is used for updating high-dimensional parameter vectors using INLA \citep{johannesson2022approximate} or MCMC \citep{Hazra2023}. 

We assume that the coordinates of individuals follow a non-homogeneous Poisson point process and thus model the ward-specific population counts of Bangalore using independent Poisson distributions, where we model the underlying intensity function using a Gaussian process without any sparsity-based (SPDE, for example) or low-rank approximation \citep{wikle2003hierarchical}. Following a Laplace approximation of the ward-level Poisson likelihood, we explore the Max-and-Smooth approach, where the approximate full conditional posterior of the intensity function follows a Gaussian process. Subsequently, we explore a Gibbs sampling algorithm for drawing posterior samples without involving Metropolis-Hastings steps. We discuss a careful choice of the prior distribution for the spatial range parameter of the Poisson intensity function that makes the computation and storage feasible. Further, while drawing exact samples from the underlying 786,702-dimensional multivariate normal full conditional posterior distribution for the vector of Poisson intensity parameters is infeasible, at least on a standard workstation, we discuss Monte Carlo estimates of the pixel-wise posterior means and posterior standard deviations without performing such high-dimensional draws. Implementation and architecture of coding (written in \texttt{R} and provided in the supplementary material) is crucial here to avoid memory and storage exhaustion on a standard workstation, and we also discuss these issues. In a simulation study and the Bangalore population data application, we compare our method with a non-spatial Gaussian process prior for the Poisson intensity function and with a standard Bayesian generalized linear model approach.

This paper is structured as follows: Section \ref{sec:EDA} offers a succinct overview of the Bangalore population dataset, encompassing the response variable, predictors, and potential relationships between the response and predictors. Section \ref{sec:methodology} elaborates on the methodology, model properties, and a practical approximate Bayesian computational scheme. The outcomes of a simulation study are presented in Section \ref{sec:simulation}. Moving to Section \ref{sec:application}, the proposed disaggregation method is applied to the Bangalore population dataset, with a detailed presentation of results and comparisons. Finally, Section \ref{sec:conclusion} provides concluding remarks.

\section{Bangalore population data and exploratory analysis}
\label{sec:EDA}

Our primary objective is to utilize alternative data sources as predictors, available at a resolution of 30m $\times$ 30m at 786,702 pixels, for disaggregating the ward-level population data for Bruhat Bengaluru Mahanagara Palike (BBMP), the administrative body overseeing the Bangalore metropolitan area (henceforth, Bangalore). The dataset includes ward-level information from the 2011 Census for BBMP, encompassing 198 wards. For each pixel, we have information about the specific ward that contains it, and thus we can easily map from 786,702 pixels to 198 wards and calculate the number of pixels within each ward. 

\begin{figure}[h]
\begin{center}
    \includegraphics[height=0.35\linewidth]{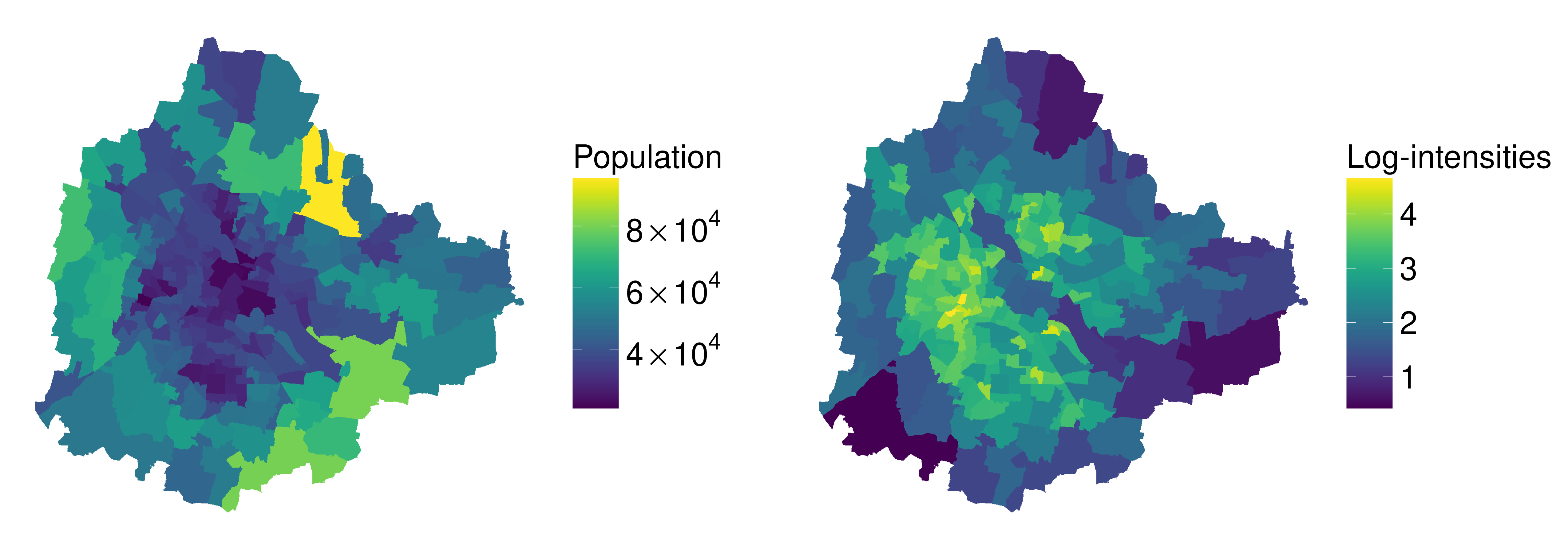}
    \caption{\label{Population}Left: Ward-level population sizes in Bangalore city area, Right: Empirical log-intensities in Bangalore.}
\end{center}
\end{figure}

The left panel of Figure \ref{Population} represents the ward-level population sizes of Bangalore. In contrast, the right panel shows empirical log-intensities (the natural logarithm of the population values divided by the respective numbers of pixels within wards). The empirical log-intensities essentially capture the logarithm of population densities across different wards. While the central region of Bangalore appears to have lower population values than the wards near the periphery, the areas of the wards in the central region are smaller. Notably, the central region, being the Information Technology hub, also known as the Silicon Valley of India, exhibits higher densities, while the suburbs near the periphery are sparsely populated. Given that the log-densities vary across different parts of the city, it is prudent to model the data using a non-homogeneous Poisson process that accommodates spatially-varying intensities across the spatial domain. This approach allows for a more realistic and nuanced modeling of the population distribution in Bangalore. 

\begin{figure}[H] %\label{fig:maps_alternative}
    \centering
    \includegraphics[height=1.25\linewidth]{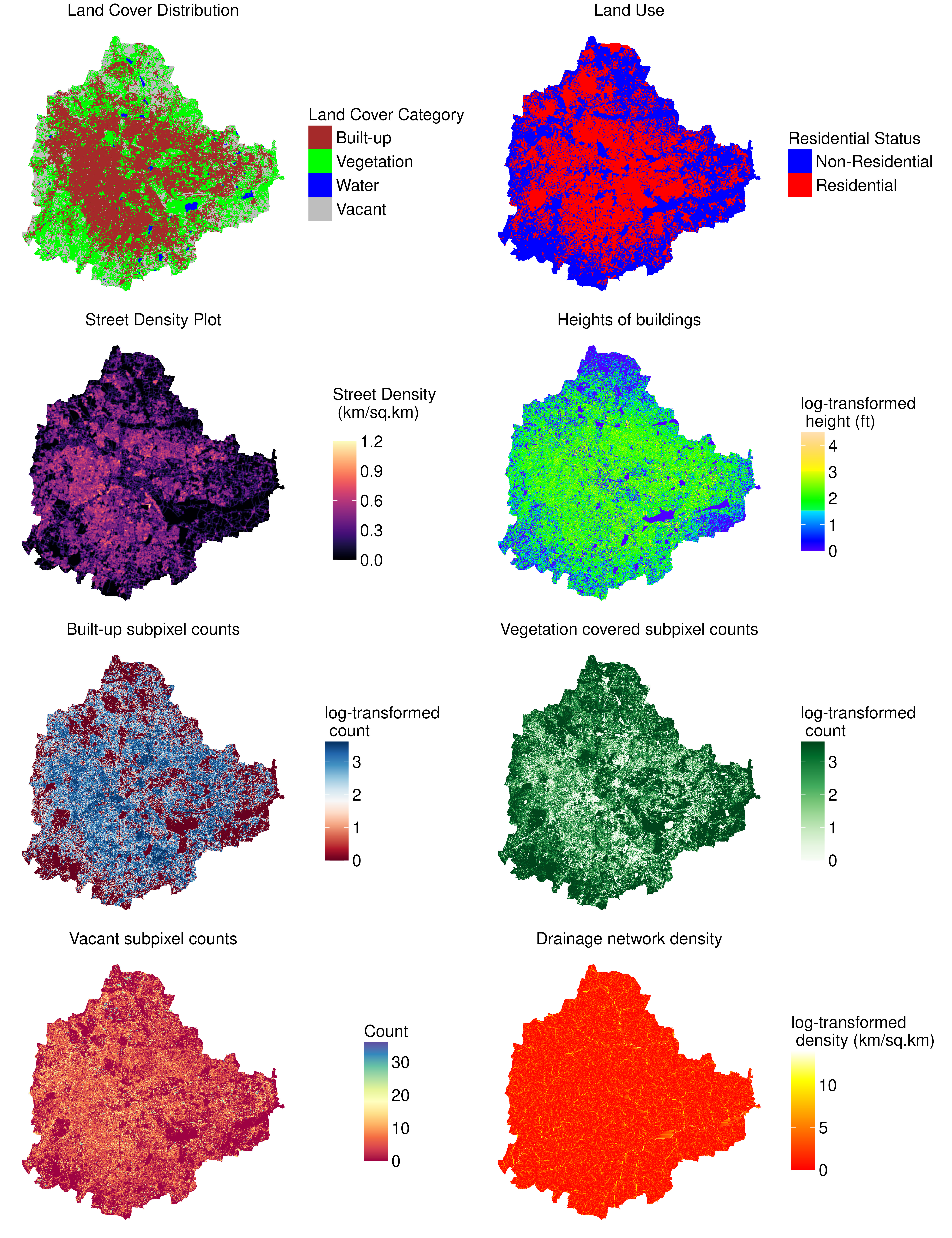}
\caption{\label{Covariates}Spatial maps of the satellite-derived alternative data sources. Here we transform some variables using the transformation $f(x)=\log(1+x)$, as otherwise, certain crucial patterns are not clearly visible. In such cases, we use the term `log-transformed'.}
\end{figure}

In addition to population data, we consider various predictors, including land cover categories (1: Built-up; 2: Vegetation; 3: Water; 4: Vacant), a binary indicator of land use (1: Residential; 0: Non-residential), street density within each pixel, building height (in meters) estimated from stereo imagery, and sub-pixel (5m × 5m) indicators denoting built-up areas, vegetation cover, or vacant land. Furthermore, we include a continuous measure of drainage network density within each pixel. We present the spatial maps of these predictors in Figure \ref{Covariates}, which are available at the pixel level (30m $\times$ 30m). We observe significant variation in the values or levels of the covariates across different regions of Bangalore, except drainage density. For instance, the central portion of the city exhibits more built-up areas, higher proportions of residential areas, taller buildings, and less vegetation compared to the suburbs. The drainage density, however, displays a unique pattern, with higher values consistently aligned with drainage lines throughout the city, irrespective of the region. This detailed analysis of covariates provides a comprehensive understanding of the diverse spatial characteristics influencing spatial population distribution and urban dynamics in Bangalore. %See Appendix B for maps displaying a comprehensive plot of the covariate layers.

\begin{figure}[H] %\label{fig:scatterplot}
    \centering
    \includegraphics[width = 1\linewidth]{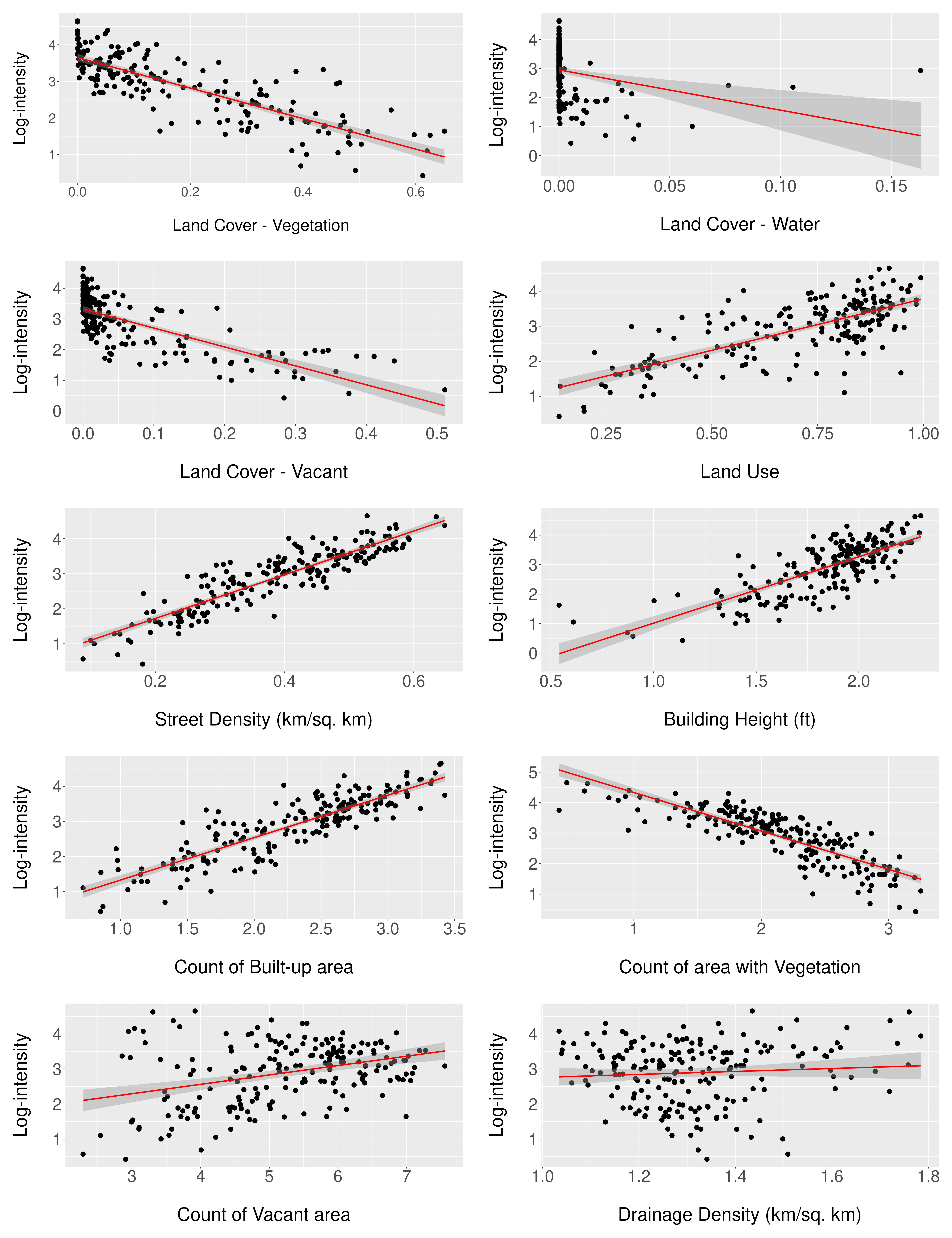}
     \caption{\label{Scatterplots}Relationships between log-intensities of population and transformed covariates in the dataset. The transformations considered here are the same as in Figure \ref{Covariates}.}
\end{figure}

We further explore the linear relationship between the response and predictors. While a nonlinear relationship is more robust and flexible, we have only 198 observations from the response variable, and hence, a linearity assumption provides parsimony. Considering a Poisson model for the data layer, a natural choice is exploring the linear relationships of the predictors with empirical log-intensities instead of actual population sizes. We present the scatterplots in Figure \ref{Scatterplots}, illustrating the relationship between the log-intensities and predictors at the ward level. The figures suggest that assuming linearity is reasonable. Here, land cover and vegetation count exhibit a negative linear relationship with the log-intensities, while the other predictors demonstrate a positive relationship. These plots justify using the available predictors for modeling the spatially-varying intensity of the Poisson point process model.

A common approach for exploring the linear relationship between the predictors and the count-type variables is the generalized linear model, specifically, a Poisson regression model with a log link. Here, for $L=198$ wards, using the \texttt{glm} function in \texttt{R}, we fit the model $Y_i \overset{\textrm{Indep}}{\sim} \text{Poisson} \left(|\mathcal{A}_i|\lambda_i\right)$, where $\log(\lambda_i) = \Tilde{\textbf{x}_i}^T \bm{\beta}$ for $i \in \{1,2, \ldots, L\}$, $|\mathcal{A}_i|$ denotes the number of 30m $\times$ 30m pixels within the $i$-th ward, $\Tilde{\textbf{x}_i}$ denotes the vector of predictors in Figure \ref{Scatterplots}, and $\bm{\beta}$ denotes the vector of regression coefficients. We present the results in Table \ref{GLMfit}. Except for the predictor `Land cover-Vegetation', all others exhibit a significant linear relationship indicating their importance in modeling the spatial variation of the intensity function of the underlying Poisson point model in a proper Bayesian modeling framework.

%The choice of an appropriate model is guided by certain assumptions and subsequent validation of the assumptions. The population counts can be assumed to follow the Poisson distribution, as is evident from Table \ref{GLMfit}, corresponding to a generalized linear model fit between the population and the covariates, using a log link. Mostly significant $p$-values add to the assertion.

\begin{table}[h]
\centering
\caption{\label{GLMfit} Parameter estimates, corresponding standard errors, $z$-values, and $p$-values of fitting a simple Poisson regression model to the ward-specific population sizes.} %The null values for standard errors indicate very low values and has been rounded off for representation purposes.
\begin{tabular}{lrrrr}
  \hline
 Predictor & Estimate & Std. Error & $z$-value & $p$-value \\ %Pr($>$$|$z$|$)
  \hline
Intercept & \hphantom{0}2.0037 & 0.00046 & 4314.8457 & $<$0.0001 \\ 
  Land cover-Vegetation & -0.0016 & 0.00303 & -0.5334 & 0.5938 \\ 
  Land cover-Water & -0.2472 & 0.00285 & -86.7009 & $<$0.0001 \\ 
  Land cover-Vacant & -0.3773 & 0.00219 & -172.3996 & $<$0.0001 \\ 
  Land Use & \hphantom{0}0.2616 & 0.00139 & 187.8546 & $<$0.0001 \\ 
  Street Density & \hphantom{0}0.6769 & 0.00129 & 524.6790 & $<$0.0001 \\ 
  Building Height & \hphantom{0}0.0162 & 0.00194 & 8.3462 & $<$0.0001 \\ 
  Builtup Count & -0.2661 & 0.00360 & -73.8608 & $<$0.0001 \\ 
  Vegetation Count & -0.8245 & 0.00240 & -343.5274 & $<$0.0001 \\ 
  Vacant Count & \hphantom{0}0.2426 & 0.00260 & 93.3627 & $<$0.0001 \\ 
  Drainage Density & \hphantom{0}0.0975 & 0.00462 & 21.1017 & $<$0.0001 \\ 
   \hline
\end{tabular}
\end{table}

% The choice of spatial smoothness is governed by the variogram analysis. 

\begin{figure}[h]
    \centering
    \includegraphics{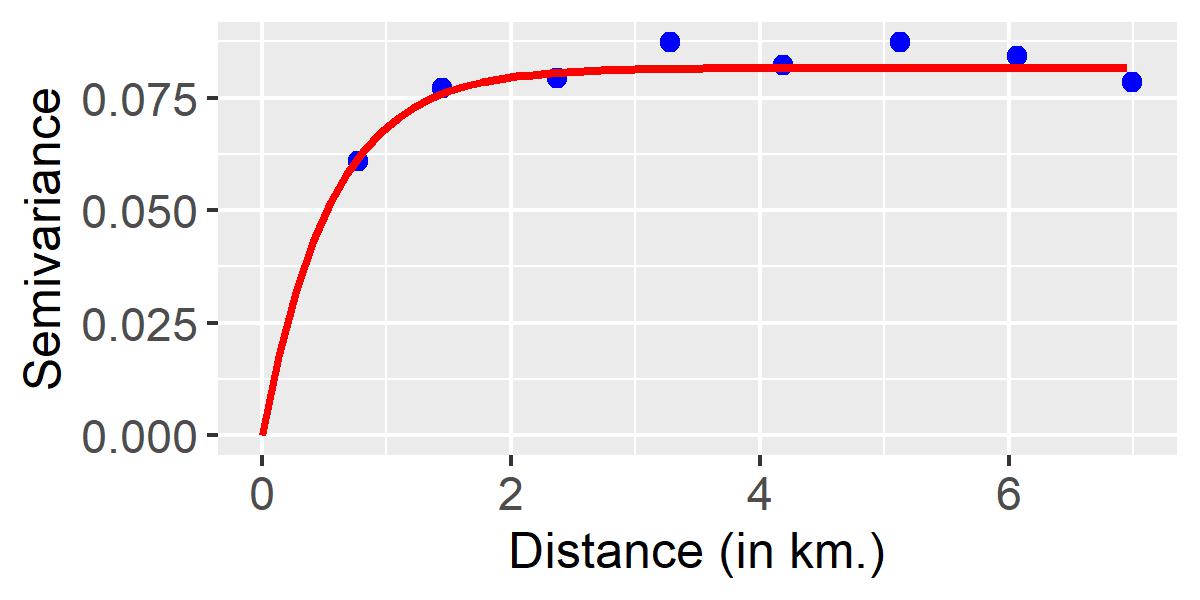}
    \caption{\label{Variogram} Empirical (dots) and fitted (line) semivariance of empirical log-intensity of population after removing effects of covariates. Here the fitted line is based on an exponential covariance kernel with parameters estimated using the function \texttt{fit.variogram} function in \texttt{R}.}
\end{figure}

While the available predictor information helps in analyzing the spatial variation, the unexplained variability needs to be modeled in a spatially-dependent framework for the Poisson intensity function. In our exploratory analysis, the empirical log-intensity is a proxy for the intensity function of the Poisson point process. We thus regress the empirical log-intensity based on the predictors shown in Figure \ref{Scatterplots} using linear regression and explore the semivariance of the residuals across the wards. Here we consider the Euclidean distance between the centroids of the wards. The semivariance of the residual process at distance $h$ is 
\begin{equation*}
    \gamma(h) = \dfrac{1}{2|N(h)|}\sum_{N(h)}(Z_i - Z_j)^2,
\end{equation*}
where $N(h)$ is the set of all pairwise distances $(i,j)$ with $i-j=h$, $|N(h)|$ being the number of unique pairs in $N(h)$, and $Z_i$ and $Z_j$ are the values of the residuals of the log-intensities at wards $i$ and $j$. We present the empirical semivariance in Figure \ref{Variogram}, where the presence of spatial correlation is clear. Further, we fit an exponential covariance kernel with parameters estimated using the function \texttt{fit.variogram} function in \texttt{R}. The corresponding fitted semivariance is presented using a solid line in Figure \ref{Variogram}, which indicates that an exponential covariance kernel provides a reasonable fit to the spatial dependence structure of the log-intensity.

%Considering an exponential kernel for the Gaussian Process covariance function, the variagram plot in Figure \ref{Variogram} indicates that there is a high spatial dependence among adjacent spatial regions, owing to a stable value of semivariance with increase in distance among the pixels. This supports the assumption of Gaussian process with exponential kernel.

% The variogram analysis also hints at the choice of covariance kernel for the Gaussian Process assumption of the smoothness term in the model. 

\section{Methodology}
\label{sec:methodology}

\subsection{Spatial disaggregation model}
\label{subsec:model}

Suppose $Y_i, i=1, \ldots, L$ denote the population sizes for $L=198$ wards in Bangalore. Each ward comprises several 30m $\times$ 30m pixels, with the number of pixels ranging between 358 and 33035, and in total, there are $P=786,702$ pixels. We further consider the area of a pixel as a single unit. Let the spatial domains for the $L$ wards be denoted by $\mathcal{A}_i \subset \mathbb{R}^2, i=1, \ldots, L$ and the entire study domain is $\mathcal{A} = \sqcup_{i=1}^L \mathcal{A}_i$. Here, $\mathcal{A}_i \cap \mathcal{A}_j = \emptyset$ for $i \neq j$. Let the vector of predictors at a spatial coordinate $\bm{s}$ be denoted by $\bm{X}(\bm{s})$ and its spatial average within the $i$-th ward is $\widetilde{\bm{X}}_i =  |\mathcal{A}_i|^{-1} \int_{\mathcal{A}_i}  \bm{X}(s) \, ds$, where $|\mathcal{A}_i|$ denotes the area of the $i$-th ward. While the predictor information is available at pixel levels, let the vector of predictors for the $j$-th pixel, with centroid $\bm{s}_j$, be denoted by $\bm{X}(\bm{s}_j)$.

We model the coordinates of the people using a non-homogeneous Poisson process (NHPP), and thus, for $\lambda_i$ denoting the average intensity within a pixel, we have 
\begin{equation} \label{eq:Y_i_def}
    \begin{split}
            Y_i &\sim \text{Poisson} \left(|\mathcal{A}_i|\lambda_i\right).
    \end{split}
\end{equation}
Given the $\lambda_i$'s are always positive and the most natural way of modeling spatial dependence is using Gaussian processes \citep[GPs,][]{gelfand2016spatial}, we model the intensity function in the log scale as
\begin{equation} \label{eq:prior_def}
    \lambda^*(\bm{s}) \equiv \log[\lambda(\bm{s})] = \bm{X}(\bm{s})^T \bm{\beta} + \eta(\bm{s}),
\end{equation}
where $\eta(\cdot)$ is a zero-mean GP with exponential covariance kernel defined over $\mathcal{A}$. We denote it by $\eta(\cdot) \sim \textrm{GP}(0, K(\cdot, \cdot))$, where $K(\bm{s},\bm{s}') = \sigma^2 \exp\{-d(\bm{s}, \bm{s}')/\phi\}$. Here, $\sigma^2$ and $\phi$ are marginal variance and spatial range parameter of the underlying GP, respectively. Given that $\mathcal{A}$ is a small geographical domain, we consider $d(\bm{s}, \bm{s}') = \Vert \bm{s} - \bm{s}'\Vert$, the Euclidean distance, which would be approximately equivalent with geodesic distance. Overall, $\lambda^*(\cdot) \sim \textrm{GP}(\bm{X}(\bm{s})^T \bm{\beta}, K(\cdot, \cdot))$ with $K(\bm{s},\bm{s}') = \sigma^2 \exp\{-\Vert \bm{s} - \bm{s}'\Vert/\phi\}$.

% Consider a spatial grid of \( P \) pixels arranged in \( N \) rows and columns. Let \(\mathbf{y}\) be the vector of unobserved values, and \(\mathbf{Y} = \left( Y_1, Y_2, \ldots, Y_L \right)^{'}\) be the vector of aggregate outcomes for \( L \) clusters or wards. Let \(\mathbf{X}\) be the matrix of scaled covariate values per pixel. The ward-level scaled covariate matrix \(\mathbf{\Tilde{X}}\) is formed by cluster-wise aggregation of \(\mathbf{X}\). K-means clustering yields \( L \) clusters as observable units, aggregating covariates accordingly. In the disaggregation model, we assume the data follow a Poisson distribution. The log-linear model has a linear predictor term for the coefficients along with a smooth spatial term that follows a smooth stochastic process (\cite{utazi2019spatial}). The Poisson intensities at the ward level are assumed to follow a nonhomogeneous Poisson process. Let \(|\mathcal{A}_i|\) represent the size of ward \(i\) and \(\lambda_i\) be the average population size in ward \(i\), commonly called the intensity function of the Poisson variable. Let \(\eta(s)\) be a smooth term following a zero-mean Gaussian process, denoted as \(\eta(.) \sim GP (0,K(.,.))\), where \(K(s,s') = \sigma^2 \exp\{-d(s,s')/\phi\}\) represents an isotropic exponential correlation function. Here, \(d(s,s')\) is the distance metric, and \(\phi\) is the range parameter.

The link between \eqref{eq:Y_i_def} and \eqref{eq:prior_def} follows from the equation
\begin{equation} \label{eq:Y_i_prior_link}
    \lambda^*_i \equiv \log[\lambda_i] = |\mathcal{A}_i|^{-1} \int_{\mathcal{A}_i}  \lambda^*(\bm{s}) \, d\bm{s} \equiv \log[\lambda(\bm{s})] = \widetilde{\bm{X}_i}^T \bm{\beta} + |\mathcal{A}_i|^{-1} \int_{\mathcal{A}_i} \eta(\bm{s}) \, d\bm{s}.
\end{equation}
Here, $\lambda_i^*$ follows a univariate normal distribution; this result follows directly from the fact that a linear combination of a multivariate normal vector is normal and integration is a linear operator. Similarly, the vector $\bm{\lambda}^* = (\lambda_1^*, \ldots, \lambda_L^*)'$ follows a multivariate normal distribution. The mean, variance, and covariances for $\lambda_i, i \in \{1, \ldots, L \}$ are given by
\begin{eqnarray} \label{eq:prior_props}
  \nonumber  \text{E}\left(\lambda^*_i\right) &=& |\mathcal{A}_i|^{-1} \int_{\mathcal{A}_i} \left[\bm{X}(\bm{s})^T\bm\beta\right] \, ds = \left[|\mathcal{A}_i|^{-1} \int_{\mathcal{A}_i}  \bm{X}(\bm{s}) \, d\bm{s}\right]^T\bm\beta = \widetilde{\bm{X}_i}^T \bm{\beta}, \\
 \nonumber  \text{Var}(\lambda_i^*) &=& \sigma^2 |\mathcal{A}_i|^{-2} \int_{\mathcal{A}_i} \int_{\mathcal{A}_i} \exp\{-\Vert \bm{s} - \bm{s}'\Vert/\phi\} \, d\bm{s} \, d\bm{s}', \\
\text{Cov}(\lambda_i^*,\lambda_j^*) &=& \sigma^2 |\mathcal{A}_i|^{-1}|\mathcal{A}_j|^{-1} \int_{\mathcal{A}_i} \int_{\mathcal{A}_j} \exp\{-\Vert \bm{s} - \bm{s}'\Vert/\phi\} \, d\bm{s} \, d\bm{s}'.
\end{eqnarray}

A natural proxy for $\widetilde{\bm{X}}_i$ is the average of $\bm{X}(\bm{s}_j)$'s with $\bm{s}_j \in \mathcal{A}_i$ and we calculate $\widetilde{\bm{X}}_i$ in the same way for further computations. Similarly, because we consider the area of a pixel as a unit, the natural proxy for $|\mathcal{A}_i|$ is the number of pixels within the $i$-th ward. The term $\text{Var}(\lambda_i^*)$ is enumerated by $\sigma^2 |\mathcal{A}_i|^{-2} \sum_{\bm{s}_j \in \mathcal{A}_i} \sum_{\bm{s}_{j'} \in \mathcal{A}_i} \exp\{-\Vert \bm{s}_j - \bm{s}_{j'} \Vert/\phi\}$. Similarly, the term $\text{Cov}(\lambda_i^*,\lambda_j^*)$ is enumerated by $\sigma^2 |\mathcal{A}_i|^{-1}|\mathcal{A}_j|^{-1} \sum_{\bm{s}_l \in \mathcal{A}_i} \sum_{\bm{s}_{l'} \in \mathcal{A}_j} \exp\{-\Vert \bm{s}_l - \bm{s}_{l'}\Vert/\phi\}$. Overall, given the parameters $\bm{\beta}$, $\sigma^2$, and $\phi$, we write $\bm\lambda^*| \bm{\beta}, \sigma^2, \phi \sim \text{N}_L(\widetilde{\bm{X}}\bm\beta, \sigma^2 \bm\Sigma_{00}^{(\phi)})$, where $\widetilde{\bm{X}}$ is obtained by stacking $\widetilde{\bm{X}}_i$'s and the $(i,j)$-th element of $\bm\Sigma_{00}^{(\phi)}$ equals 
$|\mathcal{A}_i|^{-1}|\mathcal{A}_j|^{-1} \sum_{\bm{s}_l \in \mathcal{A}_i} \sum_{\bm{s}_{l'} \in \mathcal{A}_j} \exp\{-\Vert \bm{s}_l - \bm{s}_{l'}\Vert/\phi\}$.

Our main aim in this analysis is to draw inferences about $\lambda^*(\bm{s}_j)$ for all $\bm{s}_j \in \mathcal{A}_i$ for $i=1, \ldots, L$, i.e., the vector of log-intensities at $P=786,702$ pixels. Stacking all $\lambda^*(\bm{s}_j)$'s into a vector $\bm\lambda_p^*$, the joint distribution of $\bm\lambda_p^*$ would be a $P$-dimensional multivariate normal distribution with marginal means $\text{E}[\lambda^*(\bm{s}_j)] = \bm{X}(\bm{s}_j)^T \bm{\beta}$, marginal variances $\text{Var}[\lambda^*(\bm{s}_j)] = \sigma^2$, and covariances $\text{Cov}[\lambda^*(\bm{s}_j), \lambda^*(\bm{s}_{j'})] = \sigma^2 \exp\{-\Vert \bm{s}_j - \bm{s}_{j'}\Vert/\phi\}$. Overall, we denote the joint distribution of $\bm\lambda_p^*$ by $\bm\lambda_p^*| \bm{\beta}, \sigma^2, \phi \sim \text{N}_P(\bm{X}\bm\beta, \sigma^2 \bm\Sigma_{pp}^{(\phi)})$, where $\bm{X}$ is obtained by stacking $\bm{X}(\bm{s}_j)$'s across rows and $\bm\Sigma_{pp}^{(\phi)}$ is obtained by filling its diagonal elements with ones and off-diagonal elements with $\exp\{-\Vert \bm{s}_j - \bm{s}_{j'}\Vert/\phi\}$ terms. Regarding joint distributions of $\bm\lambda_p^*$ and $\bm\lambda^*$, the vector $(\bm\lambda_p^{*T}, \bm\lambda^{*T})^T$ exhibit a joint normal distribution with a mean vector of $(\bm{X}^T | \; \widetilde{\bm{X}}^T)^T\bm\beta$ and a dispersion matrix of $\sigma^2 [(\bm{\Sigma}_{pp}^{(\phi)} \; \bm{\Sigma}_{p0}^{(\phi)}), (\bm{\Sigma}_{0p}^{(\phi)} \; \bm{\Sigma}_{00}^{(\phi)})]$. The elements of the cross-covariance matrix $\sigma^2 \bm{\Sigma}_{p0}^{(\phi)}$ are filled with terms $\text{Cov}[\lambda^*(\bm{s}_j), \lambda^*_i] = |\mathcal{A}_i|^{-1} \sum_{\bm{s}_l \in \mathcal{A}_i} \exp\{-\Vert \bm{s}_j - \bm{s}_l \Vert/\phi\}$. Thus,
\begin{equation} \label{eq:post_pred}
    \bm\lambda_p^* | \bm\lambda^*,\bm\beta,\sigma^2,\phi \sim \text{N}_P\left(\bm{X}\bm\beta + \bm{\Sigma}_{p0}^{(\phi)}\bm{\Sigma}_{00}^{(\phi)-1} \left(\bm\lambda^* - \widetilde{\bm{X}}\bm\beta\right), \sigma^2 \left(\bm{\Sigma}_{pp}^{(\phi)}-\bm{\Sigma}_{p0}^{(\phi)}\bm{\Sigma}_{00}^{(\phi)-1}\bm{\Sigma}_{0p}^{(\phi)}\right)\right).
\end{equation}

We further specify hyperpriors for $\bm{\beta}$, $\sigma^2$, and $\phi$. Assuming there are $m$ predictors available ($m=10$ in our data application) except the intercept, we assume $\bm\beta \sim \text{N}_{m+1}\left(\textbf{0},100^2 I_{m+1}\right)$ which is a weakly-informative conjugate prior. By similar logic, for $\sigma^2$, we choose the hyperprior $\sigma^2 \sim \text{Inverse-Gamma}\left(0.01,0.01\right)$. The choice of hyperprior for $\phi$ is crucial as the computational burden heavily depends on the hyperprior selection. Based on evidence from the exploratory analysis, we notice that the range parameter is likely to be between 75 meters and 575 meters, and considering 30m (distance between two first-order neighboring pixels) as a unit, $\phi$ is likely to vary between 2.5 and 17.5. We then choose the prior $\phi \sim \text{Dicrete-Uniform}\{2.5,2.75,3, \ldots, 17.5\}$. Although we guess the possible range of values of $\phi$ based on exploratory analysis, it is not an empirical Bayes approach. Ideally, the lower limit and upper limit can be tuned to allow a wider range of values, along with a finer grid of values, but the computation can be infeasible; we discuss further details about the computational bottleneck in Section \ref{subsec:computation}.

Finally, we write the overall hierarchical Bayesian model as 
\begin{eqnarray} \label{eq:hierarchical}
 \nonumber  && Y_i|\lambda_i^* \overset{\textrm{Indep}}{\sim} \text{Poisson} \left(|\mathcal{A}_i|\exp\{\lambda^*_i\}\right), \\
 \nonumber  && \bm\lambda^* = (\lambda^*_1, \ldots, \lambda^*_L)' | \bm\beta, \sigma^2, \phi \sim \text{N}_L\left(\bm{\widetilde{X}}\bm\beta, \sigma^2 \bm\Sigma_{00}^{(\phi)}\right), \\
 \nonumber && \bm\lambda_p^* | \bm\lambda^*,\bm\beta,\sigma^2,\phi \sim \text{N}_P\left(\bm{X}\bm\beta + \bm{\Sigma}_{p0}^{(\phi)}\bm{\Sigma}_{00}^{(\phi)-1} \left(\bm\lambda^* - \widetilde{\bm{X}}\bm\beta\right), \sigma^2 \left(\bm{\Sigma}_{pp}^{(\phi)}-\bm{\Sigma}_{p0}^{(\phi)}\bm{\Sigma}_{00}^{(\phi)-1}\bm{\Sigma}_{0p}^{(\phi)}\right)\right) \\
 \nonumber && \bm\beta \sim \text{N}_{m+1}\left(\textbf{0},100^2 \bm{I}_{m+1}\right),\\
 \nonumber && \sigma^2 \sim \text{Inverse-Gamma}\left(0.01,0.01\right),\\
 && \phi \sim \text{Dicrete-Uniform}\{2.5,2.75,3, \ldots, 17.5\}.
\end{eqnarray}

%We further discuss the model properties and computational details in the next subsections.

% The Poisson spatial model is articulated as follows:
% \begin{equation} \label{eq:lambda_i_def}
%     \begin{split}
%             Y_i &\sim \text{Poisson} \left(|\mathcal{A}_i|\lambda_i\right), \\
%             \log(\lambda_i) &= \Tilde{\mathbf{x}_i}^T \bm{\beta} + |\mathcal{A}_i|^{-1} \int_{\mathcal{A}_i} \eta(s) \, ds \quad \forall i \in \{1,2, \ldots, L\}.
%     \end{split}
% \end{equation}
% We introduce \(\lambda_i^* = \log(\lambda_i)\) for \( i \in \{1,2,\ldots,L\} \). The spatially varying log-population density is expressed as:
% \begin{equation} \label{eq:prior_def}
%     \log(\lambda(s)) = \mathbf{x(s)}^T \bm{\beta} + \eta(s)
% \end{equation}
% for the \( s^{th} \) spatial location (pixel), where \( s \in \{1,2,\ldots,P\} \). We introduce \(\lambda_p^*(s) = \log(\lambda(s))\) for \( s \in \{1,2,\ldots,P\} \). The aim to predict the intensities or log-intensities $\bm\lambda_p^*$ for all the pixels. 

\subsection{Model properties}
\label{subsec:properties}

% Considering a Gaussian process prior $\lambda^*(\cdot)  \sim \text{GP}\left(\bm{X}(\cdot)^T\bm\beta, K(\cdot,\cdot)\right)$ leads to the mean, variance, and covariance structures for $\lambda_i, i \in \{1, \ldots, L \}$ as follows.
% \begin{equation} \label{eq:prior_props}
%     \begin{split}
%         \text{E}\left(\lambda^*_i\right) &= |\mathcal{A}_i|^{-1} \int_{\mathcal{A}_i} \left[\bm{X}(\bm{s})^T\bm\beta\right] \, ds = \left[|\mathcal{A}_i|^{-1} \int_{\mathcal{A}_i}  \bm{X}(\bm{s}) \, d\bm{s}\right]^T\bm\beta = \widetilde{\bm{X}_i}^T \bm{\beta}, \\
%         \text{Var}(\lambda_i^*) &= |\mathcal{A}_i|^{-2} \int_{\mathcal{A}_i} \int_{\mathcal{A}_i} \sigma^2 \exp\{-d(\bm{s},\bm{s}')/\phi\} \, d\bm{s} \, d\bm{s}', \\
%         \text{Cov}(\lambda_i^*,\lambda_j^*) &= |\mathcal{A}_i|^{-1}|\mathcal{A}_j|^{-1} \int_{\mathcal{A}_i} \int_{\mathcal{A}_j} \sigma^2 \exp\{-d(\bm{s},\bm{s}')/\phi\} \, d\bm{s} \, d\bm{s}'. \\
%     \end{split}
% \end{equation}
% Here the term $\text{Cov}(\lambda_i^*,\lambda_j^*)$ is clearly a monotonically increasing function of $\phi$ and thus a one-to-one relationship between the spatial range of the log-intensity function $\lambda^*(\cdot)$ and $\lambda_i^*$ exists. Besides, $\text{Var}(\lambda_i^*)$ is also an increasing function of $\phi$. In \eqref{eq:model_def}, we redefine $\lambda_i$ as $\lambda_i = \exp\{\lambda_i^*\}$ to transform it into a linear model in \eqref{eq:prior_def}.  % $\textbf{x}_i^T \bm{\beta}$ and dispersion $|\mathcal{A}_i|^{-2} \int_{\mathcal{A}_i} \int_{\mathcal{A}_i} \sigma^2 \exp\{-d(s,s')/\phi\} \, ds \, ds'$.

After integrating out the latent Gaussian process $\lambda^*(\cdot)$, the marginal mean, variance, and covariances of $Y_i, i \in \{1, \ldots, L \}$ are as follows. %Assuming data independence and utilizing the given statistical summaries, we derive:
\begin{eqnarray}
 \nonumber   \text{E}(Y_i) &=& |\mathcal{A}_i| \cdot \exp\{\widetilde{\bm{X}_i}^T \bm{\beta}+\psi\},\\
\nonumber    \text{Var}(Y_i) &=& |\mathcal{A}_i|^2 \cdot \exp\left[2\widetilde{\bm{X}}_i^T \bm{\beta}+2\psi\right]\left(\exp\{\psi\}-1\right) + |\mathcal{A}_i| \cdot \exp\{\widetilde{\bm{X}_i}^T \bm{\beta}+\psi\}, \\ 
 \nonumber       && ~~~~~~\text{where } \psi = \frac{\sigma^2}{2}|\mathcal{A}_i|^{-2} \int_{\mathcal{A}_i} \int_{\mathcal{A}_i} \exp\{-d(\bm{s},\bm{s}')/\phi\} \, d\bm{s} \, d\bm{s}',\\
        \text{Cov}(Y_i,Y_j) &=& \text{E}(Y_iY_j) - \text{E}(Y_i) \cdot \text{E}(Y_j) \neq 0, \text{ as } \text{Cov}(\lambda_i^*,\lambda_j^*) \neq 0.
\end{eqnarray}

Keeping other parameters fixed, the regression term $\widetilde{\bm{X}_i}^T \bm{\beta}$ appears both in the first and second order moments of $Y_i$'s and its larger value implies larger mean and variance terms. Similarly, increasing $\sigma^2$ and/or $\phi$ implies larger means. A higher value of $\phi$ indicates increased spatial association and thus increased spatial aggregates on an average. A Poisson distribution with a fixed intensity parameter is equidispersed and it is always overdispersed for a random intensity parameter. Thus, we also have $\text{Var}(Y_i) > \text{E}(Y_i)$ for $i=1, \ldots, L$.

% Y_i &\sim \text{Poisson} \left(|\mathcal{A}_i|e^{\lambda_i^*}\right) \quad \forall i \in \{1,2, \ldots, L\} \\

\subsection{Computation}
\label{subsec:computation}

The first major challenge lies in the computation of the $\bm{\Sigma}_{00}^{(\phi)}$ matrices and both computation and storage of the $\bm{\Sigma}_{p0}^{(\phi)}$ matrices. For each value of $\phi$, we need to calculate the $198 \times 198$ dimensional symmetric matrix $\bm{\Sigma}_{00}^{(\phi)}$ and thus it involves calculating 19,701 unique pair of elements. As mentioned in Section \ref{subsec:model}, the $(i,j)$-th element of $\bm{\Sigma}_{00}^{(\phi)}$ is calculated by $|\mathcal{A}_i|^{-1}|\mathcal{A}_j|^{-1} \sum_{\bm{s}_l \in \mathcal{A}_i} \sum_{\bm{s}_{l'} \in \mathcal{A}_j} \exp\{-\Vert \bm{s}_l - \bm{s}_{l'}\Vert/\phi\}$, where the number of pixels varies between 358 and 33035 across wards and thus involves huge computational burden. Given the computation for 19,701 pairs of elements can be done in parallel, we perform parallel computing across 20 cores of a workstation equipped with AMD Ryzen 9 processor with 64 GB DDR4 RAM and 4 TB storage, and the overall time for calculating $\bm{\Sigma}_{00}^{(\phi)}$ is approximately 30 minutes. Each $\bm{\Sigma}_{00}^{\phi}$ matrix required a storage space of 288 KB. We use the same computational and storage facilities for subsequent calculations. Further, we again calculate the elements of the $786,702 \times 198$-dimensional $\bm{\Sigma}_{p0}^{(\phi)}$ matrix in parallel across 20 cores, and the computation time is approximately 36 minutes. Besides, it requires a storage cost of 1.2 GB. 

Overall, if the parameter space of $\phi$ involves $n_\phi$ elements, the total computation time is approximately $66 \times n_\phi$ minutes, and it requires a storage cost of approximately $1.2 \times n_\phi$ GB. Naturally, a continuous prior for $\phi$ is naturally infeasible. As a result, we choose a discrete uniform prior for $\phi$ in Section \ref{subsec:model} along with a motivation of keeping $n_\phi$ moderate yet exploring as many values of $\phi$ as possible where the posterior mass is distributed. Finally, we choose the prior $\phi \sim \text{Dicrete-Uniform}\{2.5,2.75,3, \ldots, 17.5\}$ where $n_\phi = 61$ elements. The trace plot of $\phi$ in Figure \ref{dataMCMC} shows that the choice of our prior is reasonable. While the conditional distribution of $\bm{\lambda}_p^*$ in \eqref{eq:post_pred} involves $\bm{\Sigma}_{pp}^{(\phi)}$ matrices as well, we mainly focus on the marginal distributions of the elements of $\bm{\lambda}_p^*$ which involve only the diagonal entries of $\bm{\Sigma}_{pp}^{(\phi)}$ and they are equal to one. Hence, we never calculate the $786,702 \times 786,702$-dimensional $\bm{\Sigma}_{pp}^{(\phi)}$ matrices, which is neither computationally feasible nor in terms of storage.

% the computational procedure in this study faced significant time and space constraints. Let \( n_\phi \) denote the number of choices of \( \phi \) considered. Each \(\bm{\Sigma}_{00}^{\phi}\) matrix required a storage space of 288 KB, resulting in a total storage requirement of \( n_\phi \times 288 \) KB, which is manageable. However, the time constraint was noteworthy, with the total time required for calculating \( n_\phi \) such matrices being \( n_\phi \times 30 \) minutes. In contrast, due to the large size of \(\mathbf{P}\) (786,702), the computation of \(\bm{\Sigma}_{p0}^{(\phi)}\) matrices faced both time and space constraints. The total space required to store these matrices was \( n_\phi \times 1.2 \) GB, and the total computation time was \( n_\phi \times 36 \) minutes. These computations were performed on a device equipped with 64 GB of RAM and an AMD Ryzen processor, utilizing 20 cores to distribute the computational load. Despite the substantial hardware resources, the computational challenge was significant due to the intensive demands of both time and space.

Due to our non-homogeneous Poisson process assumption in Section \ref{subsec:model}, the population values at both the ward and pixel levels are count data that follow a Poisson distribution. Consequently, the likelihood for $\lambda_i^*$'s is non-Gaussian. Given that each ward comprises a large number of pixels and the total population count of a ward is the sum of the pixel-level population sizes, the sampling distribution of the maximum likelihood estimator of $\lambda_i^*$, i.e., $\widehat{\lambda}_i^* \left( = \log(Y_i / |\mathcal{A}_i|) \right)$ would be approximately normal, by employing the Lindeberg-Feller Central Limit Theorem. This basically provides the Laplace approximation that allows us to handle the non-Gaussian likelihood effectively. Consequently, $\widehat{\lambda}_i^*$ follows a (approximately) normal distribution with mean $\lambda_i^*$ and variance $I(\widehat{\lambda}_i^*)^{-1}$. Here, $I(\widehat{\lambda}_i^*) = Y_i$, thus yielding:
\begin{equation}
    L(\lambda_i^*) \propto \exp\left[-\dfrac{1}{2}\left(\lambda_i^*-\widehat{\lambda_i^*}\right)^2Y_i \right].
\end{equation}

Here, $\widehat{\lambda}_i^*$'s are empirical log-intensities shown in the right panel of Figure \ref{Population}. Denoting the vector of ward-wise population sizes by $\bm{Y} = (Y_1, \ldots, Y_L)'$, the vector of empirical log-intensities by $\widehat{\bm{\lambda}} = (\widehat{\lambda}_1^*, \ldots, \widehat{\lambda}_L^*)$, and a diagonal matrix with its diagonal elements equal to $Y_i^{-1}, i=1, \ldots, L$ by $\textrm{diag}(\bm{1} / \bm{Y})$, we write $\widehat{\bm{\lambda}} | \bm{\lambda} \sim \textrm{N}_L(\bm{\lambda}, \textrm{diag}(\bm{1} / \bm{Y}))$. We later use the notation $\textrm{diag}(\bm{Y})$ to denote a diagonal matrix with its diagonal elements equal to $Y_i, i=1, \ldots, L$. As a result of the above Laplace approximation, the first layer of the Bayesian hierarchical model in \eqref{eq:hierarchical}, i.e., $Y_i|\lambda_i^* \overset{\textrm{Indep}}{\sim} \text{Poisson} \left(|\mathcal{A}_i|\exp\{\lambda^*_i\}\right)$ can be replaced with $\widehat{\bm{\lambda}} | \bm{\lambda} \sim \textrm{N}_L(\bm{\lambda}, \textrm{diag}(\bm{1} / \bm{Y}))$. The idea of replacing the non-Gaussian likelihood in a hierarchical Bayesian model with a Gaussian likelihood based on maximum likelihood and further performing Markov chain Monte Carlo (MCMC) or Integrated Nested Laplace Approximation (INLA) to smooth the underlying latent Gaussian process specification of the spatially-varying parameters have been explored in the literature and called a `max-and-smooth' approach \citep{hrafnkelsson2021max}. This Gaussian approximation of the likelihood allows us to circumvent Metropolis-Hastings sampling, enabling us to proceed with the Gibbs sampling \citep{gelfand2000gibbs} while maintaining conjugacy. 

% For this investigation, we assumed that $\phi$ was adjusted using a trial-and-error approach. 
% In this study, our dataset comprises the random variables $Y_1, Y_2,\ldots, Y_L$, where $L$ represents the number of clusters. The parameters within our model consist of $\lambda_1^*,\lambda_2^*,\ldots,\lambda_L^*$, while the hyperparameters include $\bm\beta$, $\sigma^2$, and $\phi$. The priors for our parameters are specified as follows:
% \begin{equation}
%     \begin{split}
%          \bm\lambda^*| \bm\beta, \sigma^2 &\sim \text{N}_L\left(\bm{\Tilde{X}}\bm\beta, \sigma^2 \bm\Sigma_{00}^{(\phi)}\right) \\
%          \bm\beta &\sim \text{N}_{m+1}\left(\textbf{0},100^2 I_{m+1}\right)\\
%          \sigma^2 &\sim \text{Inverse-Gamma}\left(0.01,0.01\right)\\
%          \bm\phi &\sim \text{Dicrete-Uniform}\{2.5,2.75,3, \ldots, 17.5\}\\
%     \end{split}
% \end{equation}

% \begin{table}[h]
% \centering
% \begin{tabular}{rrrr}
%   \hline
%  & Setup & Coverage & DIC \\ 
%   \hline
% 1 &   $\lambda^* \sim N(\Tilde{X}\beta, \sigma^2\Sigma_{00})$ & 1.00 & -1708.44 \\ 
%   2 &   $\lambda^* \sim N(\Tilde{X}\beta, \sigma^2 I)$ & 0.94 & 2275.25 \\ 
%   3 &   $\lambda^* = \Tilde{X}\beta$ & 1.00 & 6687.93 \\ 
%   4 &   BayesGLM (w/o smoothness) & 0.93 & 6656.24 \\ 
%   5 &   BayesGLM (with smoothness) & 0.12 & 42469.67 \\ 
%    \hline
% \end{tabular}
% \end{table}

%Here, $\bm{\Tilde{X}}$ represents the averaged covariate matrix corresponding to the clusters, and $\bm\Sigma_{00}^{(\phi)} = ((\sigma_{ij})) = |\mathcal{A}_i|^{-1}|\mathcal{A}_j|^{-1} \sum_{\mathcal{A}_i} \sum_{\mathcal{A}_j} \exp\{-d(x,y)/\phi\}$. 

In Gibbs sampling with Laplace approximation, we need to draw samples from the posterior $\pi(\bm{\lambda}^*, \bm{\lambda}^*_p, \bm{\beta}, \sigma^2, \phi | \hat{\bm{\lambda}}^*) = \pi(\bm{\lambda}^*_p | \bm{\lambda}^*, \bm{\beta}, \sigma^2, \phi, \hat{\bm{\lambda}}^*) \times \pi(\bm{\lambda}^*, \bm{\beta}, \sigma^2, \phi | \hat{\bm{\lambda}}^*)$. The first term on the right side does not depend on $\hat{\bm{\lambda}}^*$ and is given by \eqref{eq:post_pred}. We thus focus on drawing samples from $\pi(\bm{\lambda}^*, \bm{\beta}, \sigma^2, \phi | \hat{\bm{\lambda}}^*)$. Here, in a Gibbs sampling framework, we draw samples from $\pi(\bm{\lambda}^*| \bm{\beta}, \sigma^2, \phi, \hat{\bm{\lambda}}^*)$, $\pi(\bm{\beta} | \bm{\lambda}^*, \sigma^2, \phi, \hat{\bm{\lambda}}^*)$, $\pi(\sigma^2 | \bm{\lambda}^*, \bm{\beta}, \phi , \hat{\bm{\lambda}}^*)$, and $\pi(\phi | \bm{\lambda}^*, \bm{\beta}, \sigma^2, \hat{\bm{\lambda}}^*)$. As $\vert \mathcal{A}_i \vert$ are known and fixed, given $\hat{\bm{\lambda}}^*$, we know $\bm{Y}$ and vice versa. The first three full conditional posteriors are given by
\begin{eqnarray}
\nonumber && \bm\lambda^*| \bm\beta, \sigma^2, \phi, \hat{\bm{\lambda}}^* \sim \text{N}_L\left(\bm\mu^*,\bm\Sigma^*\right), \\
\nonumber && \bm{\beta} | \bm{\lambda}^*, \sigma^2, \phi, \hat{\bm{\lambda}}^* \equiv  \bm{\beta} | \bm{\lambda}^*, \sigma^2, \phi \sim \text{N}_{m+1}\left(\bm\mu_1,\bm\Sigma_1\right),\\
&& \sigma^2 | \bm{\lambda}^*, \bm{\beta}, \phi , \hat{\bm{\lambda}}^* \equiv \sigma^2 | \bm{\lambda}^*, \bm{\beta}, \phi \sim \text{Inverse-Gamma}\left(A,B\right),
\end{eqnarray}
where the expressions for $\bm\mu^*$, $\bm\Sigma^*$, $\bm\mu_1$, $\bm\Sigma_1$, $A$, and $B$ are given by
\[ \bm\Sigma^* = \left(\dfrac{1}{\sigma^2}\bm\Sigma_{00}^{(\phi)-1} + \text{diag}(\bm{Y})\right)^{-1}, \qquad \bm\mu^* = \bm\Sigma^*\left(\dfrac{1}{\sigma^2}\bm\Sigma_{00}^{(\phi)-1}\widetilde{\bm{X}}\bm\beta + \text{diag}(\bm{Y})\widehat{\bm\lambda}^*\right),\]
\[ \bm\Sigma_1 = \left(\dfrac{1}{\sigma^2}\widetilde{\bm{X}}^T\bm\Sigma_{00}^{(\phi)-1}\widetilde{\bm{X}} + \dfrac{1}{100^2} \bm{I}_{m+1}\right)^{-1}, \qquad \bm\mu_1 = \bm\Sigma_1\left(\dfrac{1}{\sigma^2}\widetilde{\bm{X}}^T\bm\Sigma_{00}^{(\phi)-1}\bm\lambda^*\right),\]
\[ A = 0.01 + \dfrac{L}{2}, \qquad B = 0.01 + \dfrac{1}{2}\left(\bm\lambda^* - \widetilde{\bm{X}}\bm\beta\right)^T\bm\Sigma_{00}^{(\phi)-1}\left(\bm\lambda^* - \widetilde{\bm{X}}\bm\beta\right).\]

Further, $\pi(\phi | \bm{\lambda}^*, \bm{\beta}, \sigma^2, \hat{\bm{\lambda}}^*)$ is equal to $\pi(\phi | \bm{\lambda}^*, \bm{\beta}, \sigma^2) \propto f_{\textrm{N}_L}(\bm{\lambda}^*; \bm{\widetilde{X}}\bm\beta, \sigma^2 \bm\Sigma_{00}^{(\phi)})$, where $f_{\textrm{N}_L}(\cdot; \widetilde{\bm{\mu}}, \widetilde{\bm{\Sigma}})$ denotes the dentisy of a $L$-variate normal distribution with mean vector and covariance matrix given by $\widetilde{\bm{\mu}}$ and $\widetilde{\bm{\Sigma}}$, respectively. Thus, we draw samples from $\pi(\phi | \bm{\lambda}^*, \bm{\beta}, \sigma^2)$ using a probability proportional to size (PPS) sampling on values of $\phi$, where the weights assigned to the values in the parameter space of $\phi$ are $\pi(\phi | \bm{\lambda}^*, \bm{\beta}, \sigma^2)$ scaled to ensure they sum to one. Finally, we obtain MCMC chains for the hyperparameters $\bm{\lambda}^*, \bm{\beta}, \sigma^2$, and $\phi$. 

Given that drawing samples from the 786,702-dimensional multivariate normal distribution $\pi(\bm{\lambda}^*_p | \bm{\lambda}^*, \bm{\beta}, \sigma^2, \phi, \hat{\bm{\lambda}}^*)$ is an infeasible problem, we focus on obtaining the quantities \(\text{E}(\bm{\lambda}_p^*|\bm{Y})\) and the vector of marginal posterior standard deviations \(\text{SD}(\bm{\lambda}_p^*|\bm{Y})\) which are essential. To achieve this, we employ the concepts of conditional mean and conditional covariance to derive the mean and standard deviation from the posterior predictive distribution, as outlined below:
\begin{eqnarray}
 \nonumber \text{E}(\bm{\lambda}_p^*|\bm{Y}) &=& \text{E}\left[\text{E}(\bm{\lambda}_p^* | \bm{\lambda}^*, \bm{\beta}, \sigma^2, \phi, \bm{Y})|\bm{Y}\right] \\ 
\nonumber   &=& \text{E}\left[\bm{X}\bm{\beta} +        \bm{\Sigma}_{p0}^{(\phi)}\bm{\Sigma}_{00}^{(\phi)-1} \left(\bm{\lambda}^* - \widetilde{\bm{X}}\bm{\beta}\right)|\bm{Y}\right] \\ 
        &\approx& \dfrac{1}{\text{B}}\sum_{b=1}^{\text{B}} \left[\bm{X}\bm{\beta}^{(b)} + \bm{\Sigma}_{p0}^{(\phi^{(b)})}\bm{\Sigma}_{00}^{(\phi^{(b)})-1} \left(\bm{\lambda}^{*(b)} - \widetilde{\bm{X}}\bm{\beta}^{(b)}\right)\right].
\end{eqnarray}
% \begin{align}
%     \begin{split}
%         \text{E}(\bm{\lambda}_p^*|\bm{Y}) &= \text{E}\left[\text{E}(\bm{\lambda}_p^* | \bm{\lambda}^*, \bm{\beta}, \sigma^2, \phi, \bm{Y})|\bm{Y}\right] \\ 
%         &= \text{E}\left[\bm{X}\bm{\beta} + \bm{\Sigma}_{p0}^{(\phi)}\bm{\Sigma}_{00}^{(\phi)-1} \left(\bm{\lambda}^* - \widetilde{\bm{X}}\bm{\beta}\right)|\bm{Y}\right] \\ 
%         &\approx \dfrac{1}{\text{B}}\sum_{b=1}^{\text{B}} \left[\bm{X}\bm{\beta}^{(b)} + \bm{\Sigma}_{p0}^{(\phi^{(b)})}\bm{\Sigma}_{00}^{(\phi^{(b)})-1} \left(\bm{\lambda}^{*(b)} - \widetilde{\bm{X}}\bm{\beta}^{(b)}\right)\right].
%     \end{split}
% \end{align}

\begin{eqnarray}
\nonumber     \text{Cov}(\bm{\lambda}_p^*|\bm{Y}) &=& \text{E}\left[\text{Cov}(\bm{\lambda}_p^* | \bm{\lambda}^*, \bm{\beta}, \sigma^2, \phi, \bm{Y})|\bm{Y}\right] + \text{Cov}\left[\text{E}(\bm{\lambda}_p^* | \bm{\lambda}^*, \bm{\beta}, \sigma^2, \phi, \bm{Y})|\bm{Y}\right]\\ 
\nonumber     &=& \text{E}\left[\sigma^2 \left(\bm{\Sigma}_{pp}^{(\phi)} - \bm{\Sigma}_{p0}^{(\phi)}\bm{\Sigma}_{00}^{(\phi)-1}\bm{\Sigma}_{0p}^{(\phi)}\right)|\bm{Y}\right]\\ 
\nonumber && + \text{Cov}\left[\bm{X}\bm{\beta} + \bm{\Sigma}_{p0}^{(\phi)}\bm{\Sigma}_{00}^{(\phi)-1} \left(\bm{\lambda}^* - \widetilde{\bm{X}}\bm{\beta}\right)|\bm{Y}\right] \\
\nonumber    &\approx & \dfrac{1}{\text{B}}\sum_{i=1}^{\text{B}} \left[\sigma^{2(b)} \left(\bm{\Sigma}_{pp}^{(\phi^{(b)})} - \bm{\Sigma}_{p0}^{(\phi^{(b)})}\bm{\Sigma}_{00}^{(\phi^{(b)})-1}\bm{\Sigma}_{0p}^{(\phi^{(b)})}\right)|\bm{Y}\right] \\ 
        && + \text{Sample Cov}\left[\left\lbrace\bm{X}\bm{\beta}^{(b)} + \bm{\Sigma}_{p0}^{(\phi^{(b)})}\bm{\Sigma}_{00}^{(\phi^{(b)})-1} \left(\bm{\lambda}^{*(b)} - \widetilde{\bm{X}}\bm{\beta}^{(b)}\right) \right\rbrace_{b=1}^B\right].
\end{eqnarray}

While computing the terms like $\bm{X}\bm{\beta}^{(b)}$, it is advisable not to form matrices of dimension $786,702 \times B$ where $B$ is large in general; for our application, we have $B=1500$ and the storage cost of a $786,702 \times 1500$-dimensional matrix is approximately 9 GB making the computation vulnerable on a standard desktop. We divide the computation into different wards and calculate such terms for one ward at a time. Besides, we focus only on the diagonal elements of $\text{Cov}(\bm{\lambda}_p^*|\bm{Y})$, which involves the diagonal elements of $\bm{\Sigma}_{pp}^{(\phi^{(b)})}$ only and they are ones. % by our model assumption.

\section{Simulation Study}
\label{sec:simulation}

The purpose of this simulation study is to compare and contrast the proposed model with a standard parametric Bayesian generalized linear model that does not incorporate a Laplace approximation, a parametric Bayesian multiple linear regression model that incorporates a Laplace approximation, and a semiparametric Bayesian model that incorporates a Laplace approximation but replaces the dense Gaussian process prior for the intensity function with spatially-independent white noise. The covariates used in the study are the same as they are in the setup of the problem statement. We simulate data under three different situations as follows. Here each setting aims to assess the performance and robustness of the proposed model against existing and simpler models under varying degrees of spatial smoothness of the intensity function and its tiny (the coefficients of the sine and cosine terms are only 0.05 and 0.1) deviation from a linear combination of the available predictor surfaces.
\begin{eqnarray*} \label{eq:simulation_settings}
\nonumber \textrm{S1}:~~\log[\lambda(\bm{s})] &=& \bm{X}(\bm{s})^T \bm{\beta} \\
\nonumber \textrm{S2}:~~\log[\lambda(\bm{s})] &=& \bm{X}(\bm{s})^T \bm{\beta} + 0.05 \sin(2\pi s_1^*) + 0.05 \cos(2\pi s_2^*)\\
 \textrm{S3}:~~\log[\lambda(\bm{s})] &=& \bm{X}(\bm{s})^T \bm{\beta} + 0.1 \sin(2\pi s_1^*) + 0.1 \cos(2\pi s_2^*)
\end{eqnarray*}

Here, $\bm{X}(\bm{s})$ represents the predictor vector at location $\bm{s}$, and \(s_1^*\) and \(s_2^*\) are respectively the scaled row number and scaled column number of the pixel under consideration (scaled row/column number implies row/column number divided by the maximum row/column number). In all settings, the values assigned to \(\bm{\beta}\) are the least squares estimates obtained from a simple linear regression run on the empirical log-intensity and predictors described in Section~\ref{sec:EDA}. The smooth curve concatenated to the linear predictor in settings S2 and S3 adds a surface to the data that deviates from the plane spanned by $\bm{X}(\bm{s})^T \bm{\beta}$ and smooth and nonlinear to any predictor (sinusoidal) in nature. Suppose we call the true value of the vector $\bm{\lambda}^*_p$, obtained according to settings S1, S2, and S3, using a generic notation $\bm{\lambda}^{*(true)}_p$.

For each of the given settings, we analyze the results based on four different models. The first model (BayesGLM) is the Bayesian Generalized Linear Model, fit using \texttt{bayesglm} in \texttt{R}. The other models use the Laplace Approximation, approximating the Poisson likelihood with a Gaussian distribution. In the second model (Laplace), $\eta(\bm{s})$ in \eqref{eq:prior_def} is set to 0, implying no effect of the smooth term in the model. The third model (Laplace-WN) assumes that $\eta(\cdot)$ follows a Gaussian white noise process, i.e., $\eta(\bm{s}) \overset{\textrm{IID}}{\sim} \textrm{Normal}(0, \sigma^2)$. The final model (Laplace-GP), formulated in this paper, assumes that $\eta(\cdot)$ follows a Gaussian process with an exponential covariance kernel.

For each model in each setting, we report a range of statistical measures, including root mean squared error (RMSE), mean absolute deviation (MAD), average posterior standard deviation (PosSD), empirical coverage (Cover), deviance information criterion (DIC), Watanabe-Akaike information criterion (WAIC), and the computation time (in seconds). Here, RMSE and MAD denote the $L_2$ and $L_1$ distances between the posterior mean of $\bm{\lambda}^*_p$ and $\bm{\lambda}^{*(true)}_p$, respectively. Further, PosSD denotes the average of the elementwise posterior standard deviations of $\bm{\lambda}^*_p$. The measure Cover denotes the coverage of the 95\% pointwise normal approximated posterior credible regions of $\bm{\lambda}^*_p$ obtained using the element-wise posterior means and posterior standard deviations and then averaging out across the pixels. DIC and WAIC are standard measures for comparing Bayesian models. We simulate 100 datasets from each setting and fit all the four competing models and report the averages of the above-mentioned measures in Table \ref{Simulationtable}. Furthermore, in the simulation study, we consider a single value of $\phi=10$ for the model Laplace-GP wherever necessary to allow feasible computation on a standard desktop. This simplification is essential given the complexity of the entire computation process, which is constrained by both space and time limitations.

%$\text{E}() - \bm{\lambda}^{*(true)}_p$

\begin{table}[h]
\centering
\caption{\label{Simulationtable} Results of the simulation study for three data generation settings (S1, S2, and S3) and four models for each setting. We report the RMSE, MAD, posterior standard deviation, Empirical Coverage, DIC, WAIC, and computation time (in seconds). A model with smaller RMSE, MAD, PosSD, DIC, WAIC, and computation time and higher Cover is preferred.}
\begin{tabular}{lllllllll}
  \hline
Setting & Model & RMSE & MAD & PosSD & Cover & DIC & WAIC & Time(s) \\ 
  \hline
S1 & BayesGLM & 0.006 & 0.005 & 0.006 & 0.961 & 6658.068 & 2686.272 & 38.227 \\ 
  & Laplace & 0.006 & 0.005 & 0.006 & 0.961 & 6658.014 & 2686.204 & 0.414 \\ 
  & Laplace-WN & 0.007 & 0.005 & 0.079 & 1.000 & 6597.706 & 2699.187 & 48.333 \\ 
  & Laplace-GP & 0.008 & 0.006 & 0.020 & 1.000 & 6498.361 & 2738.884 & 208.213 \\
  \hline
S2 & BayesGLM & 0.074 & 0.059 & 0.006 & 0.120 & 42015.896 & 19061.157 & 38.198 \\ 
  & Laplace & 0.074 & 0.059 & 0.006 & 0.119 & 42035.586 & 19070.840 & 0.400 \\ 
  & Laplace-WN & 0.151 & 0.113 & 2.943 & 1.000 & 6563.456 & 2867.129 & 47.802 \\ 
  & Laplace-GP & 0.085 & 0.065 & 0.133 & 1.000 & 6560.979 & 2863.246 & 205.803 \\
  \hline
S3 & BayesGLM & 0.148 & 0.117 & 0.006 & 0.060 & 145714.742 & 67067.684 & 37.506 \\ 
  & Laplace & 0.148 & 0.117 & 0.006 & 0.060 & 146017.145 & 67237.061 & 0.397 \\ 
  & Laplace-WN & 0.305 & 0.229 & 5.938 & 1.000 & 6552.044 & 2859.617 & 46.714 \\ 
  & Laplace-GP & 0.171 & 0.132 & 0.265 & 1.000 & 6551.136 & 2858.162 & 202.824 \\ 
   \hline
\end{tabular}
\end{table}

%Under setting S1, where the $\eta(\bm{s})$ term in \eqref{eq:prior_def} is actually zero, the parametric models 

For all settings, the Laplace model has the least computation time, while our Laplace-GP model requires the most computation time; however, obtaining posterior estimates along with uncertainty measures for a 786,702-dimensional parameter vector in approximately 200 seconds is reasonable (apart from pre-calculated $\bm\Sigma_{00}^{(\phi)}$ and $\bm\Sigma_{p0}^{(\phi)}$ matrices). RMSE and MAD values do not vary significantly, so no model is consistently preferred based on these measures, although the Laplace-WN model performs slightly poorer in settings S2 and S3. The Laplace-WN and Laplace-GP models demonstrate significantly better empirical coverage than the others across all settings, particularly in S2 and S3, where there exists a smooth surface in the data generating mechanism. For S1, DIC and WAIC values are almost similar across the different models. However, in S2 and S3, the Laplace-GP model is favored, as it has the lowest DIC and WAIC values—slightly lower than those of the Laplace-WN model and significantly lower than those of the BayesGLM or Laplace models. The debate remains over whether to choose the Laplace-WN or Laplace-GP model for modeling our data. This decision is guided by PosSD, which is significantly lower for the Laplace-GP model compared to the Laplace-WN model. Therefore, we prefer the Laplace-GP model after considering several measures for model comparison. Additionally, the choice of the Laplace-GP model is justified for our data as it exhibits significant spatial smoothness, as shown in the variogram study in Figure \ref{Variogram}. %Although the computation time is higher for our proposed model (termed here as Laplace-GP), we choose to proceed with it as all statistical measures of accuracy favor this model.

\section{Data application}
\label{sec:application}

\begin{figure}[h]
     \centering
     \includegraphics[width = 1\linewidth]{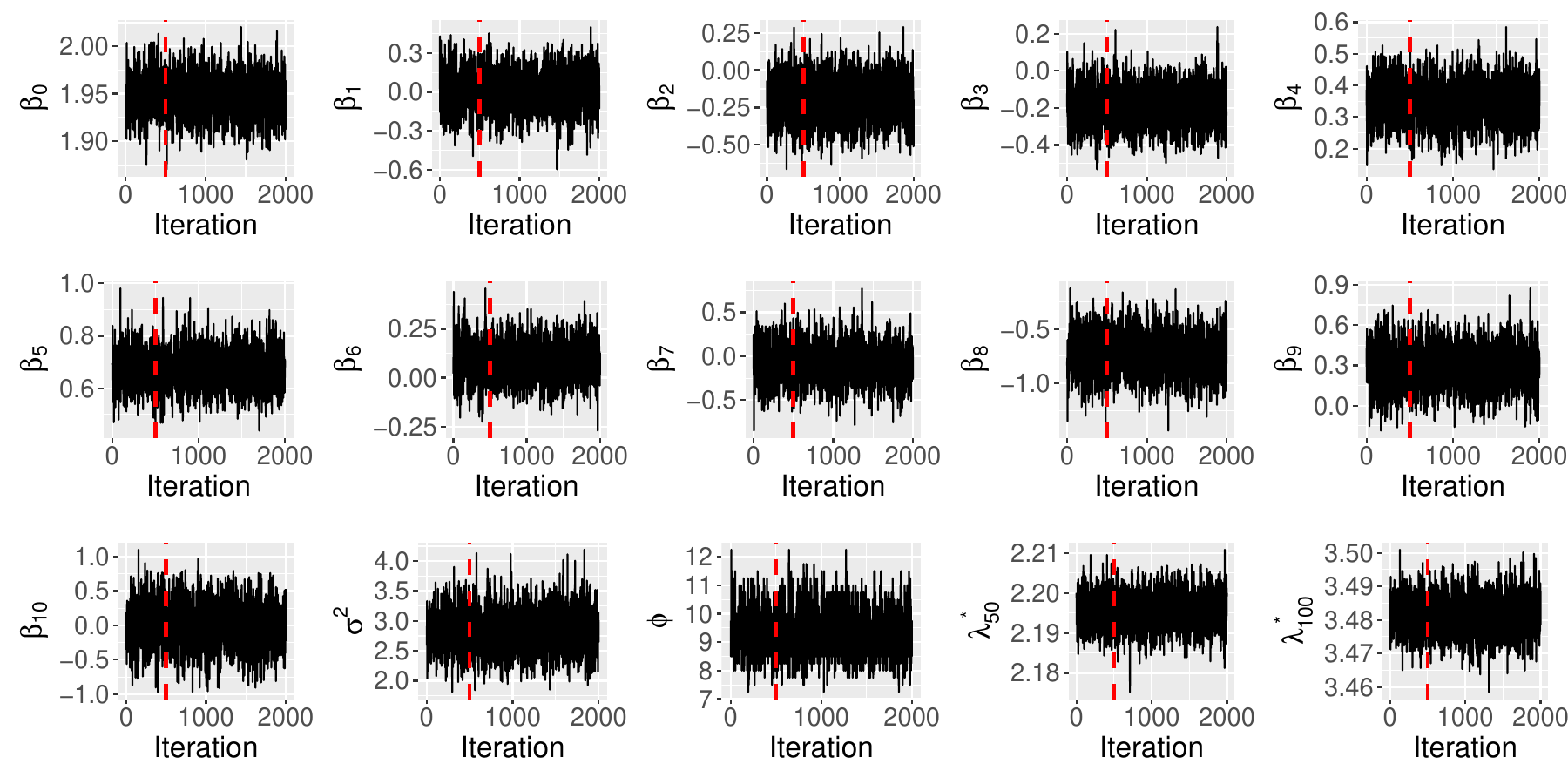}
     \caption{\label{dataMCMC}Trace plots for the model hyperparameters $\beta$'s, $\sigma^2$, and $\phi$, and also for $\lambda^*_{50}$ and $\lambda^*_{100}$. Here, the burn-in size is 500, indicated by the red vertical lines.}
\end{figure}

We apply our proposed methodology and computation scheme for the dataset described in Section \ref{sec:EDA}. To draw posterior inferences according to Section \ref{subsec:computation}, we run the MCMC chain for the hyperparameters (\(\bm{\beta}\), \(\sigma^2\), \(\phi\)) and the parameters (\(\bm{\lambda}^*\)). The process is carried out with 500 burn-in samples and 1500 posterior samples. The trace plots of the MCMC chains for all hyperparameters are illustrated using trace plots in Figure \ref{dataMCMC}. We also show the trace plots of the 50-th and 100-th (arbitrarily chosen) elements of $\bm{\lambda}^*$. These plots demonstrate reasonable mixing and convergence of the MCMC chains for all parameters and hyperparameters. Furthermore, we observe that the chains exhibit fast convergence for all the hyperparameters and parameters. For $\bm{\lambda}^*$, the initial value is chosen as the vector of empirical log-intensities $\hat{\bm{\lambda}}^*$. Further, we regress $\hat{\bm{\lambda}}^*$ on $\widetilde{\bm{X}}$ in a simple linear regression format and start the chains of $\bm{\beta}$ and $\sigma^2$ from the estimated regression coefficients and the average of the squared residuals. Given the reasonable quality of the MCMC chains, it is valid to proceed with posterior inferences based on these samples.

\begin{table}[h]
\centering
\caption{\label{coefficient}Posterior means, standard deviations (SD), and equal-tailed 95\% credible intervals of the model hyperparameters. The names of predictors corresponding to $\beta$'s are provided within brackets in the first column. }
\begin{tabular}{lcccc}
  \hline
Parameter & Posterior Mean & Posterior SD & Credible Interval \\ 
  \hline
$\beta_0$ (Intercept) & 1.9469 & 0.0212 & (1.9049, 1.9882) \\ 
  $\beta_1$ (Land cover-Vegetation) & 0.0082 & 0.1480 & (-0.2770, 0.3055) \\ 
  $\beta_2$ (Land cover-Water) & -0.1970 & 0.1449 & (-0.4830, 0.0788) \\ 
  $\beta_3$ (Land cover-Vacant) & -0.1918 & 0.1061 & (-0.3953, 0.0179) \\ 
  $\beta_4$ (Land Use) & 0.3431 & 0.0681 & (0.2074, 0.4758) \\ 
  $\beta_5$ (Street Density) & 0.6730 & 0.0737 & (0.5337, 0.8164) \\ 
  $\beta_6$ (Building Height) & 0.0861 & 0.0931 & (-0.0943, 0.2684) \\ 
  $\beta_7$ (Builtup Count) & -0.0872 & 0.2086 & (-0.4968, 0.3193) \\ 
  $\beta_8$ (Vegetation Count) & -0.7476 & 0.1846 & (-1.0986, -0.3691) \\ 
  $\beta_9$ (Vacant Count) & 0.2812 & 0.1534 & (-0.0124, 0.5851) \\ 
  $\beta_{10}$ (Drainage Density) & 0.0068 & 0.3139 & (-0.6262, 0.6232) \\ 
  $\sigma^2$ & 2.7865 & 0.3566 & (2.1562, 3.5380) \\ 
  $\phi$ & 9.2992 & 0.7863 & (8, 11) \\ 
   \hline
\end{tabular}
\end{table}

Table \ref{coefficient} presents the posterior inference for the hyperparameters. The posterior means of the elements of $\bm{\beta}$ are positive for Land cover-Vegetation, Land cover-Vacant, Builtup Count, and Vegetation count; except for Builtup Count, a negative posterior mean for the other three is reasonable as the population density is likely to be less in areas with vegetation. Considering Builtup Count, the 95\% posterior credible interval includes zero indicating that the corresponding coefficient is not significantly negative, which is reasonable. A parametric Bayesian generalized linear model (\texttt{bayesglm}) provides a significant negative effect of Builtup Count in Table \ref{GLMfit}, which is counter-intuitive. Unlike \texttt{bayesglm}, the 95\% credible intervals based on the proposed semiparametric Bayesian spatial model include zero for several elements of $\bm{\beta}$ correctly identifying the significantly linearly-related predictors.

\begin{figure}[h]
\begin{center}
    \includegraphics[height=0.35\linewidth]{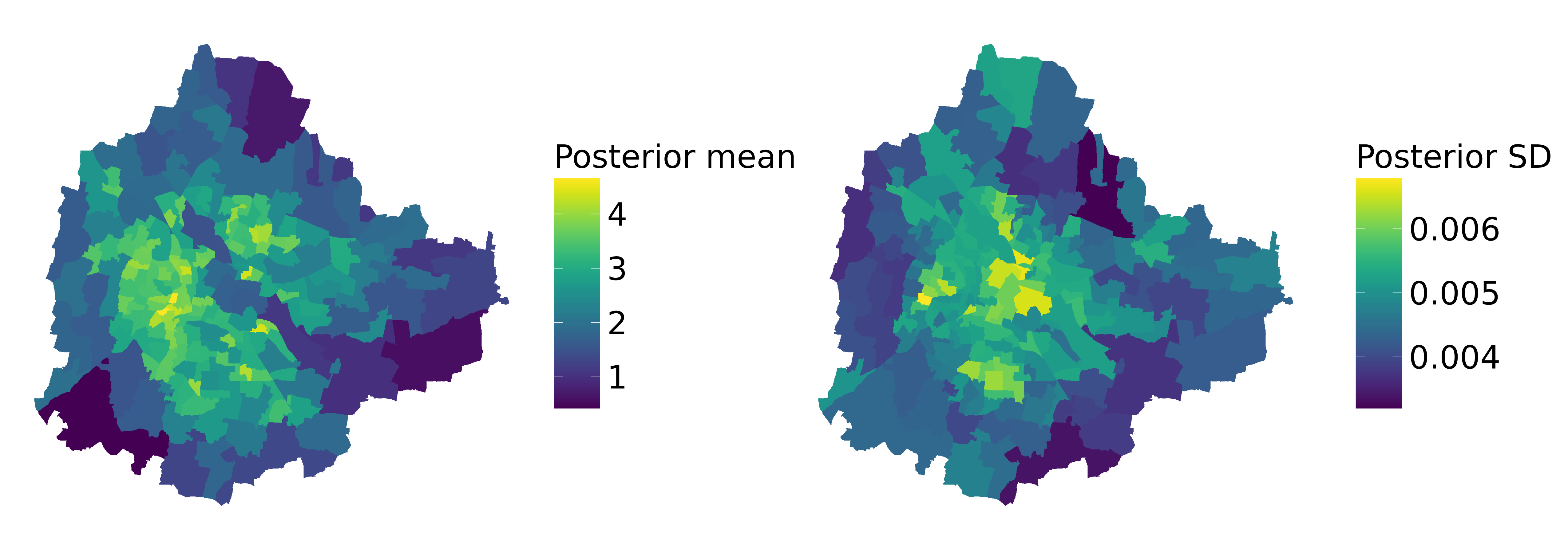}
    \caption{\label{LambdaInference}Left: Ward-level posterior means of $\lambda^*_i$'s, Right: Ward-level posterior standard deviations for $\lambda_i^*$'s.}
\end{center}
\end{figure}

Figure \ref{LambdaInference} illustrates the posterior mean and standard deviation for the parameter $\bm{\lambda}^*$ that we map onto the actual outline of Bangalore.  Higher values of the posterior mean are observed in the center of the city compared to the outskirts. This indicates that ward-level population densities are actually higher in the central areas of Bangalore than in the wards far from the center. The posterior standard deviation is small throughout the city, although the central areas exhibit slightly higher variability in the estimates than the peripheral wards. Further, it is observed that the estimated intensities closely mirror the empirical intensities, hinting that the max-and-smooth approach in Section \ref{subsec:computation} provides limited spatial smoothness in the posterior distribution of $\lambda^*$ but does not change abruptly, which is reasonable.

% Consequently, the density estimates in the central regions can vary more than those near the boundary. 

% By bypassing complicated integrals through a judicious use of conditional expectation and variance, we are left with posterior predictive mean and standard deviation for the pixel-level rates. 
% As regards the mean, it is observed that there are a few negative values, which may not seem appropriate given that the obtained Bayesian estimates are rates for some Poisson distribution.
% The dark blue dots near the east-central part and the south-west part are two such instances. However, this anomaly can be attributed to the use of an approximation to the Poisson likelihood by a normal distribution to aid the use of the Gibbs sampler. 

\begin{figure}[h]
\begin{center}
    \includegraphics[height=0.35\linewidth]{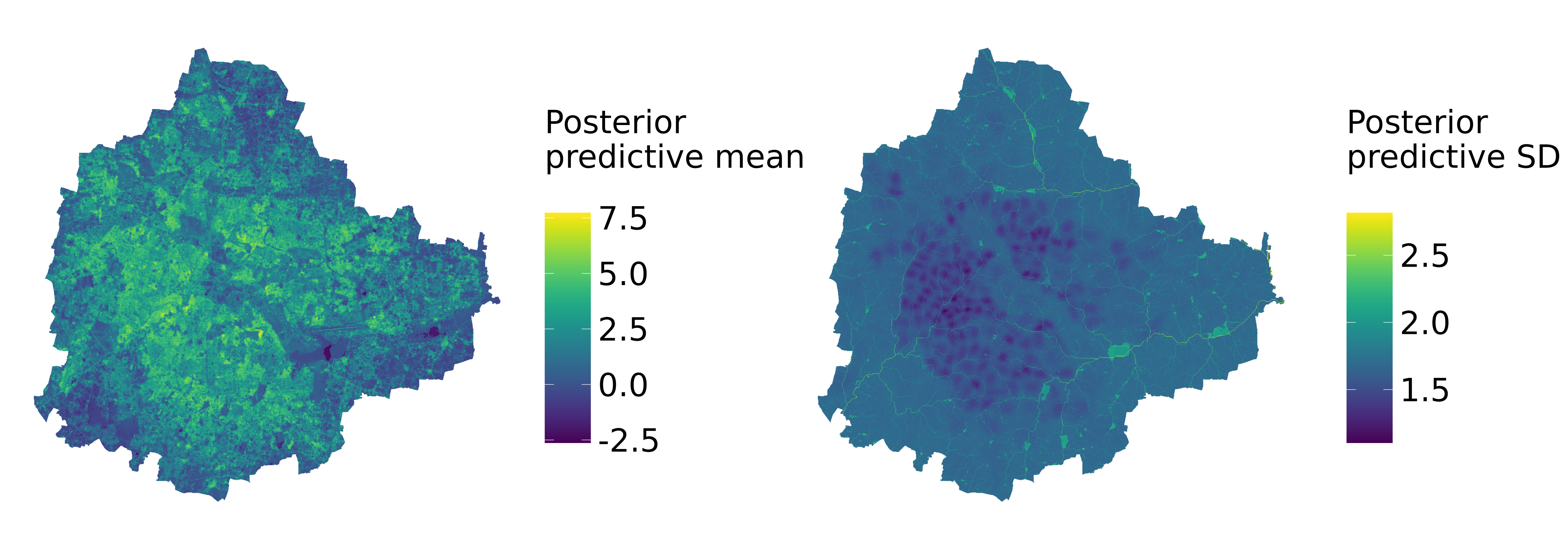}
    \caption{\label{PixelLambdaInference}Left: Posterior predictive means for $\bm{\lambda}_p^*$, i.e., the disaggregated population density estimates (in log scale). Right: Posterior predictive standard deviation for $\bm{\lambda}_p^*$, providing the uncertainty estimates in estimating disaggregated population densities (in log scale).}
\end{center}
\end{figure}

In the final stage of computation, our goal is to obtain inference about the marginal posterior means and standard deviations of the elements of $\bm{\lambda}^*_p$. These rates provide important insights into the density of the population at a much higher resolution (here, 30m $\times$ 30m) compared to the ward-level data we have. Hence, the problem of spatial downscaling has been addressed, and we now have the rates at the pixel level, which can be used for redefining boundaries of wards and provide important ideas about social and ecological aspects like water distribution networks. Figure \ref{PixelLambdaInference} gives us a summary of the above, where we can see the spatial variability of the disaggregated population density estimates (in log scale), along with the underlying uncertainty estimates, across the city. Overall, the map structure shows some similarity to the ward-level maps, in the sense that the wards with higher intensities also have pixels with higher intensities, as we observe the yellow regions towards the center. The standard deviation map does not show much variation, with the exception of several blue dots in the central region. These hint at lower variability due to smaller ward definitions towards the center, leading to more precise posterior inferences near the center.

\section{Conclusion}
\label{sec:conclusion}

The Census of India provides publicly available datasets at the ward level that are large portions of a metropolitan area, and wards are generally not based on precise information about public services like water distribution networks. As a result, for public services and several environmental factors, having a precise idea of population density at a finer spatial resolution is necessary. Motivated by this problem, we discuss a semiparametric hierarchical Bayesian latent Gaussian model framework that uses alternative data sources to obtain precise estimates of a high-resolution population density map. While existing literature discusses similar spatial disaggregation problems \citep{JSSv106i11}, the implementation assumes model-based approximations of the underlying Gaussian process, using a stochastic partial differential equation, for example \citep{lindgren2011explicit}. Besides, usual implementations generally do not involve disaggregating the available data into almost a million pixels, with a few exceptions like \cite{utazi2019spatial}. We discuss tricks for computing covariance matrices for the aggregates on a limited computation resource, the max-and-smooth approach of \cite{hrafnkelsson2021max} in our setup, and also how to obtain elementwise posterior means and standard deviations. The simulation study ensures a better performance of the proposed methodology over existing and simplified models.

The results indicate that the significant predictors of population density are the variables: Land Use, Street Density, Vegetation Count, where the last one is negatively related. After considering the effects of all the available predictors, the estimated spatial range ($3\phi$ for exponential correlation kernel) is approximately 837 meters, which is realistic for fast-developing cities like Bangalore, where sharp spatial variation between rural and urban areas are generally visible due to lower land price of the rural areas compared to the central regions of the city. The Bangalore city often faces a massive water crisis \citep{Fareha2024}. The impact on the environment can also be assessed at a high spatial resolution based on our findings. 

While our methodology is motivated by the Bangalore population dataset, it can be applied to various disaggregation problems in various scientific disciplines. While we explore the Poisson likelihood setup, similar ideas can be used for any general non-Gaussian likelihood. Motivated by the exploratory data analysis, we stick to the exponential covariance; however, other kernels like squared exponential and Mat\'ern can also be used in different setups.

%Throughout this article, our focus has been on leveraging ward-level aggregated population sizes from Census of India datasets to estimate these variables at the 30m $\times$ 30m pixel level within each respective ward. Employing a Bayesian framework, we aimed to derive pixel-level estimates. The conducted simulation study assessed the effectiveness of employing Markov Chain Monte Carlo (MCMC) methods for estimation, yielding promising results and suggesting potential applications in ward and regional reconstruction where new population units are required.

%The population study conducted on Bangalore demonstrated favorable outcomes, with empirical and estimated intensities exhibiting close similarity. Despite relying on the Gibbs Sampler for our Bayesian inference without utilizing tools like Integrated Nested Laplace Approximations (INLA), our study holds relevance for disaggregation modeling. The simulation study establishes the superiority of our proposed model over several similar settings and models. However, the extensive work that ensued due to the large dataset size posed challenges and led us to use approximations like the Laplace approximation. The constraints with respect to space and time were significant, which posed difficulties in following the usual steps in the procedure. Nevertheless, the results are robust enough to be applicable in real-world domains. While our final predictions proved satisfactory, our posterior estimates leave room for improvement, a topic we intend to explore in future research.

\newpage

\baselineskip=14pt
\bibliographystyle{apalike}

\bibliography{PHRB_2024}

\begin{thebibliography}{}

\bibitem[Alber and Piégay, 2011]{ALBER2011343}
Alber, A. and Piégay, H. (2011).
\newblock {Spatial disaggregation and aggregation procedures for characterizing fluvial features at the network-scale: Application to the Rhone basin (France)}.
\newblock {\em Geomorphology}, 125(3):343--360.

\bibitem[Anjoy et~al., 2019]{anjoy2019estimation}
Anjoy, P., Chandra, H., and Basak, P. (2019).
\newblock {Estimation of disaggregate-level poverty incidence in Odisha under area-level hierarchical Bayes small area model}.
\newblock {\em Social Indicators Research}, 144:251--273.

\bibitem[Arambepola et~al., 2022]{arambepola2022simulation}
Arambepola, R., Lucas, T.~C., Nandi, A.~K., Gething, P.~W., and Cameron, E. (2022).
\newblock A simulation study of disaggregation regression for spatial disease mapping.
\newblock {\em Statistics in Medicine}, 41(1):1--16.

\bibitem[Balakrishnan, 2020]{balakrishnan2020method}
Balakrishnan, K. (2020).
\newblock A method for urban population density prediction at 30m resolution.
\newblock {\em Cartography and Geographic Information Science}, 47(3):193--213.

\bibitem[Berliner et~al., 2000]{berliner2000long}
Berliner, L.~M., Wikle, C.~K., and Cressie, N. (2000).
\newblock {Long-lead prediction of Pacific SSTs via Bayesian dynamic modeling}.
\newblock {\em Journal of climate}, 13(22):3953--3968.

\bibitem[Brooks et~al., 2011]{brooks2011handbook}
Brooks, S., Gelman, A., Jones, G., and Meng, X.-L. (2011).
\newblock {\em {Handbook of Markov chain Monte Carlo}}.
\newblock CRC press, New York.

\bibitem[Bullock et~al., 2023]{bullock2023latent}
Bullock, Z., Zimmaro, P., Lavrentiadis, G., Wang, P., Ojomo, O., Asimaki, D., Rathje, E.~M., and Stewart, J.~P. (2023).
\newblock {A latent Gaussian process model for the spatial distribution of liquefaction manifestation}.
\newblock {\em Earthquake Spectra}, 39(2):1189--1213.

\bibitem[Cisneros et~al., 2023]{cisneros2023combined}
Cisneros, D., Gong, Y., Yadav, R., Hazra, A., and Huser, R. (2023).
\newblock A combined statistical and machine learning approach for spatial prediction of extreme wildfire frequencies and sizes.
\newblock {\em Extremes}, 26(2):301--330.

\bibitem[de~Oliveira et~al., 2023]{de2023collaborative}
de~Oliveira, G.~A., da~Silva~Ribeiro, A.~A., and Cirilo, J.~A. (2023).
\newblock {Collaborative spatial information as an alternative data source for hydrodynamic model calibration: a Pernambuco State case study, Brazil}.
\newblock {\em Natural Hazards}, pages 1--25.

\bibitem[Earnest et~al., 2010]{earnest2010small}
Earnest, A., Beard, J.~R., Morgan, G., Lincoln, D., Summerhayes, R., Donoghue, D., Dunn, T., Muscatello, D., and Mengersen, K. (2010).
\newblock {Small area estimation of sparse disease counts using shared component models-application to birth defect registry data in New South Wales, Australia}.
\newblock {\em Health \& place}, 16(4):684--693.

\bibitem[Foulkes and Newbold, 2008]{foulkes2008using}
Foulkes, M. and Newbold, K.~B. (2008).
\newblock {Using alternative data sources to study rural migration: examples from Illinois}.
\newblock {\em Population, Space and Place}, 14(3):177--188.

\bibitem[Gelfand, 2000]{gelfand2000gibbs}
Gelfand, A.~E. (2000).
\newblock Gibbs sampling.
\newblock {\em Journal of the American statistical Association}, 95(452):1300--1304.

\bibitem[Gelfand and Schliep, 2016]{gelfand2016spatial}
Gelfand, A.~E. and Schliep, E.~M. (2016).
\newblock Spatial statistics and gaussian processes: A beautiful marriage.
\newblock {\em Spatial Statistics}, pages 86--104.

\bibitem[Golder and Macy, 2011]{golder2011diurnal}
Golder, S.~A. and Macy, M.~W. (2011).
\newblock Diurnal and seasonal mood vary with work, sleep, and daylength across diverse cultures.
\newblock {\em Science}, 333(6051):1878--1881.

\bibitem[Harva et~al., 2008]{harva2008algorithms}
Harva, M. et~al. (2008).
\newblock {\em Algorithms for approximate Bayesian inference with applications to astronomical data analysis}.
\newblock Teknillinen korkeakoulu.

\bibitem[Hazra et~al., 2021]{hazra2021realistic}
Hazra, A., Huser, R., and Bolin, D. (2021).
\newblock Realistic and fast modeling of spatial extremes over large geographical domains.
\newblock {\em arXiv preprint arXiv:2112.10248}.

\bibitem[Hazra et~al., 2023]{Hazra2023}
Hazra, A., Huser, R., and J{\'o}hannesson, {\'A}.~V. (2023).
\newblock {\em Bayesian Latent Gaussian Models for High-Dimensional Spatial Extremes}, pages 219--251.
\newblock Springer International Publishing, Cham.

\bibitem[Hrafnkelsson and Bakka, 2023]{Hrafnkelsson2023}
Hrafnkelsson, B. and Bakka, H. (2023).
\newblock {\em Bayesian Latent Gaussian Models}, pages 1--80.
\newblock Springer International Publishing, Cham.

\bibitem[Hrafnkelsson et~al., 2021]{hrafnkelsson2021max}
Hrafnkelsson, B., Siegert, S., Huser, R., Bakka, H., and J{\'o}hannesson, {\'A}.~V. (2021).
\newblock {Max-and-Smooth: a two-step approach for approximate Bayesian inference in latent Gaussian models}.
\newblock {\em Bayesian Analysis}, 16(2):611--638.

\bibitem[Irekponor et~al., 2022]{irekponor2022framework}
Irekponor, V., Abdul-Rahman, M., Agunbiade, M., and Bustamente, A. (2022).
\newblock A framework to determine micro-level population figures using spatially disaggregated population estimates.
\newblock {\em arXiv preprint arXiv:2212.02020}.

\bibitem[J{\'o}hannesson et~al., 2022]{johannesson2022approximate}
J{\'o}hannesson, {\'A}.~V., Siegert, S., Huser, R., Bakka, H., and Hrafnkelsson, B. (2022).
\newblock {Approximate Bayesian inference for analysis of spatiotemporal flood frequency data}.
\newblock {\em The Annals of Applied Statistics}, 16(2):905--935.

\bibitem[Li et~al., 2023]{li2023combined}
Li, C.-H., Mao, J.-J., Wu, Y.-J., Zhang, B., Zhuang, X., Qin, G., and Liu, H.-M. (2023).
\newblock {Combined impacts of environmental and socioeconomic covariates on HFMD risk in China: A spatiotemporal heterogeneous perspective}.
\newblock {\em PLOS Neglected Tropical Diseases}, 17(5):e0011286.

\bibitem[Lindgren et~al., 2011]{lindgren2011explicit}
Lindgren, F., Rue, H., and Lindstr{\"o}m, J. (2011).
\newblock {An explicit link between Gaussian fields and Gaussian Markov random fields: the stochastic partial differential equation approach}.
\newblock {\em Journal of the Royal Statistical Society Series B: Statistical Methodology}, 73(4):423--498.

\bibitem[Machado et~al., 2021]{machado2021alternative}
Machado, A.~M., Giehl, E. L.~H., Fernandes, L.~P., Ingram, S.~N., and Daura-Jorge, F.~G. (2021).
\newblock Alternative data sources can fill the gaps in data-poor fisheries.
\newblock {\em ICES Journal of Marine Science}, 78(5):1663--1671.

\bibitem[Martino et~al., 2011]{martino2011approximate}
Martino, S., Akerkar, R., and Rue, H. (2011).
\newblock {Approximate Bayesian inference for survival models}.
\newblock {\em Scandinavian Journal of Statistics}, 38(3):514--528.

\bibitem[Mertens and Lambin, 1997]{mertens1997spatial}
Mertens, B. and Lambin, E.~F. (1997).
\newblock {Spatial modelling of deforestation in southern Cameroon: spatial disaggregation of diverse deforestation processes}.
\newblock {\em Applied Geography}, 17(2):143--162.

\bibitem[Monteiro et~al., 2019]{ijgi8080327}
Monteiro, J., Martins, B., Murrieta-Flores, P., and Pires, J.~M. (2019).
\newblock Spatial disaggregation of historical census data leveraging multiple sources of ancillary information.
\newblock {\em ISPRS International Journal of Geo-Information}, 8(8).

\bibitem[Muhling et~al., 2018]{muhling2018potential}
Muhling, B.~A., Gait{\'a}n, C.~F., Stock, C.~A., Saba, V.~S., Tommasi, D., and Dixon, K.~W. (2018).
\newblock Potential salinity and temperature futures for the chesapeake bay using a statistical downscaling spatial disaggregation framework.
\newblock {\em Estuaries and Coasts}, 41:349--372.

\bibitem[Murphy et~al., 2023]{murphy2023bayesian}
Murphy, K.~J., Ciuti, S., Burkitt, T., and Morera-Pujol, V. (2023).
\newblock Bayesian areal disaggregation regression to predict wildlife distribution and relative density with low-resolution data.
\newblock {\em Ecological Applications}, 33(8):e2924.

\bibitem[Naaz, 2024]{Fareha2024}
Naaz, F. (2024).
\newblock Bengaluru water supply cut alert! bswwb warns of water shortage on these days.

\bibitem[Nandi et~al., 2023]{JSSv106i11}
Nandi, A.~K., Lucas, T. C.~D., Arambepola, R., Gething, P., and Weiss, D.~J. (2023).
\newblock {disaggregation: An R Package for Bayesian Spatial Disaggregation Modeling}.
\newblock {\em Journal of Statistical Software}, 106(11):1–19.

\bibitem[Paige et~al., 2022]{paige2022spatial}
Paige, J., Fuglstad, G.-A., Riebler, A., and Wakefield, J. (2022).
\newblock Spatial aggregation with respect to a population distribution: Impact on inference.
\newblock {\em Spatial Statistics}, 52:100714.

\bibitem[Pollet et~al., 2015]{pollet2015taking}
Pollet, T.~V., Stulp, G., Henzi, S.~P., and Barrett, L. (2015).
\newblock Taking the aggravation out of data aggregation: A conceptual guide to dealing with statistical issues related to the pooling of individual-level observational data.
\newblock {\em American journal of primatology}, 77(7):727--740.

\bibitem[Ponciano et~al., 2009]{ponciano2009hierarchical}
Ponciano, J.~M., Taper, M.~L., Dennis, B., and Lele, S.~R. (2009).
\newblock Hierarchical models in ecology: confidence intervals, hypothesis testing, and model selection using data cloning.
\newblock {\em Ecology}, 90(2):356--362.

\bibitem[Roquette et~al., 2018]{roquette2018relevance}
Roquette, R., Nunes, B., and Painho, M. (2018).
\newblock {The relevance of spatial aggregation level and of applied methods in the analysis of geographical distribution of cancer mortality in mainland Portugal (2009--2013)}.
\newblock {\em Population health metrics}, 16:1--12.

\bibitem[Rudstrom et~al., 2002]{rudstrom2002data}
Rudstrom, M., Popp, M., Manning, P., and Gbur, E. (2002).
\newblock Data aggregation issues for crop yield risk analysis.
\newblock {\em Canadian Journal of Agricultural Economics/Revue canadienne d'agroeconomie}, 50(2):185--200.

\bibitem[Rue et~al., 2009]{rue2009approximate}
Rue, H., Martino, S., and Chopin, N. (2009).
\newblock {Approximate Bayesian inference for latent Gaussian models by using integrated nested Laplace approximations}.
\newblock {\em Journal of the Royal Statistical Society Series B: Statistical Methodology}, 71(2):319--392.

\bibitem[Sadik et~al., 2020]{sadik2020small}
Sadik, K., Anisa, R., and Aqmaliyah, E. (2020).
\newblock {Small Area Estimation on Zero-Inflated Data Using Frequentist and Bayesian Approach}.
\newblock {\em Journal of Modern Applied Statistical Methods}, 18(1):8.

\bibitem[Sawicki, 1973]{sawicki1973studies}
Sawicki, D.~S. (1973).
\newblock Studies of aggregated areal data: problems of statistical inference.
\newblock {\em Land Economics}, 49(1):109--114.

\bibitem[Schmid and Brown, 2000]{schmid2000bayesian}
Schmid, C.~H. and Brown, E.~N. (2000).
\newblock Bayesian hierarchical models.
\newblock {\em Methods in enzymology}, 321:305--330.

\bibitem[Segond et~al., 2007]{segond2007simulation}
Segond, M.-L., Neokleous, N., Makropoulos, C., Onof, C., and Maksimovic, C. (2007).
\newblock Simulation and spatio-temporal disaggregation of multi-site rainfall data for urban drainage applications.
\newblock {\em Hydrological sciences journal}, 52(5):917--935.

\bibitem[Shiferaw, 2023]{shiferaw2023mapping}
Shiferaw, Y.~A. (2023).
\newblock {Mapping Disaggregate-Level Agricultural Households in South Africa Using a Hierarchical Bayes Small Area Estimation Approach}.
\newblock {\em Agriculture}, 13(3):631.

\bibitem[Stumpf, 2014]{stumpf2014approximate}
Stumpf, M.~P. (2014).
\newblock {Approximate Bayesian inference for complex ecosystems}.
\newblock {\em F1000Prime Reports}, 6.

\bibitem[Sudhira et~al., 2007]{sudhira2007city}
Sudhira, H., Ramachandra, T., and Subrahmanya, M.~B. (2007).
\newblock City profile.
\newblock {\em Cities}, 24(5):379--390.

\bibitem[Tapia et~al., 2016]{tapia2016prediction}
Tapia, G., Elwany, A.~H., and Sang, H. (2016).
\newblock {Prediction of porosity in metal-based additive manufacturing using spatial Gaussian process models}.
\newblock {\em Additive Manufacturing}, 12:282--290.

\bibitem[Tasic et~al., 2016]{tasic2016applications}
Tasic, I., Porter, R.~J., and Brewer, S. (2016).
\newblock {Applications of generalized additive and Bayesian hierarchical models for areal safety analysis: case study of an urban multimodal transportation system in Chicago, Illinois}.
\newblock {\em Transportation research record}, 2601(1):99--109.

\bibitem[Tassone et~al., 2010]{tassone2010disaggregated}
Tassone, E.~C., Miranda, M.~L., and Gelfand, A.~E. (2010).
\newblock Disaggregated spatial modelling for areal unit categorical data.
\newblock {\em Journal of the Royal Statistical Society Series C: Applied Statistics}, 59(1):175--190.

\bibitem[Utazi et~al., 2019]{utazi2019spatial}
Utazi, C., Thorley, J., Alegana, V., Ferrari, M., Nilsen, K., Takahashi, S., Metcalf, C. J.~E., Lessler, J., and Tatem, A. (2019).
\newblock A spatial regression model for the disaggregation of areal unit based data to high-resolution grids with application to vaccination coverage mapping.
\newblock {\em Statistical Methods in Medical Research}, 28(10-11):3226--3241.

\bibitem[van Beurden and Douven, 1999]{van1999aggregation}
van Beurden, A.~U. and Douven, W.~J. (1999).
\newblock Aggregation issues of spatial information in environmental research.
\newblock {\em International Journal of Geographical Information Science}, 13(5):513--527.

\bibitem[Wainwright et~al., 2016]{wainwright2016hierarchical}
Wainwright, H.~M., Flores~Orozco, A., B{\"u}cker, M., Dafflon, B., Chen, J., Hubbard, S.~S., and Williams, K.~H. (2016).
\newblock {Hierarchical Bayesian method for mapping biogeochemical hot spots using induced polarization imaging}.
\newblock {\em Water Resources Research}, 52(1):533--551.

\bibitem[Wikle, 2003]{wikle2003hierarchical}
Wikle, C.~K. (2003).
\newblock Hierarchical bayesian models for predicting the spread of ecological processes.
\newblock {\em Ecology}, 84(6):1382--1394.

\bibitem[Wulder et~al., 2012]{wulder2012landsat}
Wulder, M.~A., Masek, J.~G., et~al. (2012).
\newblock Landsat legacy.
\newblock {\em Remote Sensing of Environment}, 122:1--202.

\bibitem[Yadav et~al., 2023]{yadav2023joint}
Yadav, R., Huser, R., Opitz, T., and Lombardo, L. (2023).
\newblock Joint modelling of landslide counts and sizes using spatial marked point processes with sub-asymptotic mark distributions.
\newblock {\em Journal of the Royal Statistical Society Series C: Applied Statistics}, 72(5):1139--1161.

\bibitem[Yang et~al., 2022]{yang2022classification}
Yang, Y., Gao, H., Berry, C., Carrick, D., Radjenovic, A., and Husmeier, D. (2022).
\newblock {Classification of myocardial blood flow based on dynamic contrast-enhanced magnetic resonance imaging using hierarchical Bayesian models}.
\newblock {\em Journal of the Royal Statistical Society Series C: Applied Statistics}, 71(5):1085--1115.

\bibitem[You et~al., 2009]{you2009generating}
You, L., Wood, S., and Wood-Sichra, U. (2009).
\newblock {Generating plausible crop distribution maps for Sub-Saharan Africa using a spatially disaggregated data fusion and optimization approach}.
\newblock {\em Agricultural Systems}, 99(2-3):126--140.

\end{thebibliography}

\end{document}